\documentclass[journal]{IEEEtran}

\usepackage{graphicx}  
\usepackage{url}
\usepackage{array,tabularx}
\usepackage{multicol}  
\usepackage{amsmath}
\usepackage{amsfonts}
\usepackage{cite}
\usepackage{picinpar}
\usepackage{colortbl}
\usepackage{hyperref}
\usepackage{soul}
\usepackage{csquotes}
\usepackage{pifont}
\usepackage{color}
\usepackage[english]{babel}
\usepackage{alltt}
\usepackage{enumerate}
\usepackage{dsfont}
\usepackage{siunitx}
\usepackage{breakurl}
\usepackage{epstopdf}
\usepackage{pbox}
\usepackage{mathrsfs}
\usepackage[utf8]{inputenc}
\usepackage{algorithm}
\usepackage{algpseudocode}

\newtheorem{remark}{Remark}

\ifCLASSINFOpdf
\else
\fi

\begin{document}
	\title{Deep Reinforcement Learning-based UAV Navigation and Control: A Soft Actor-Critic with Hindsight Experience Replay Approach}
	
	\author{Myoung Hoon Lee and Jun Moon
	
		\thanks{Myoung Hoon Lee is with the Research Institute of Electrical and Computer Engineering, Hanyang University, Seoul 04763, South Korea; \texttt{email:
			leemh91@hanyang.ac.kr.}}
		\thanks{Jun Moon is with the Department of Electrical Engineering, Hanyang University, Seoul 04763, South Korea; \texttt{email: junmoon@hanyang.ac.kr.}}
		\thanks{This research was supported 
		by an Institute of Information \& communications Technology Planning \& Evaluation (IITP) grant funded by the Korean government (MSIT) (No.2020-0-01373, Artificial Intelligence Graduate School Program (Hanyang University)).}}

	\markboth{Journal of \LaTeX\ Class Files,~Vol.~14, No.~8, August~2015}%
	{Shell \MakeLowercase{\textit{et al.}}: Bare Demo of IEEEtran.cls for IEEE Journals}

	\maketitle
	
	\begin{abstract}
		In this paper, we propose SACHER (soft actor-critic (SAC) with hindsight experience replay (HER)), which constitutes a class of deep reinforcement learning (DRL) algorithms. SAC is known as an off-policy model-free DRL algorithm based on the maximum entropy framework, which outperforms earlier DRL algorithms in terms of exploration, robustness and learning performance. However, in SAC, maximizing the entropy-augmented objective may degrade the optimality of learning outcomes. HER is known as a sample-efficient replay method that enhances the performance of off-policy DRL algorithms by allowing the agent to learn from both failures and successes. We apply HER to SAC and propose SACHER to improve the learning performance of SAC. More precisely, SACHER achieves the desired optimal outcomes faster and more accurately than SAC, since HER improves the sample efficiency of SAC. We apply SACHER to the navigation and control problem of unmanned aerial vehicles (UAVs), where SACHER generates the optimal navigation path of the UAV under various obstacles in operation. Specifically, we show the effectiveness of SACHER in terms of the tracking error and cumulative reward in UAV operation by comparing them with those of state-of-the-art DRL algorithms, SAC and DDPG. Note that SACHER in UAV navigation and control problems can be applied to arbitrary models of UAVs.
	\end{abstract}
	
	\begin{IEEEkeywords}
		Deep reinforcement learning, soft actor-critic, hindsight experience replay, UAV navigation and control
	\end{IEEEkeywords}
	
	\IEEEpeerreviewmaketitle
		
	\section{Introduction}\label{section1} 
	In recent years, problems of navigation and control for unmanned aerial vehicles (UAVs) have been utilized in various applications, such as wildfire monitoring, target tracking and surveillance, and formation and collision avoidance \cite{Wang_JFI_2018,Motlagh2017,Verginis2017}. Most recent studies on navigation and control of UAVs depend on model accuracy and/or prior knowledge of operation environment. However, identifying accurate model and operation environment information is challenging due to the lack of complete environment information. Deep reinforcement learning (DRL) would be an alternative approach to overcoming such limitations, since DRL does not need the UAV model information, which can be applied to various operation environments \cite{Salt2020, Shao2019, Li2020}. In addition, the model-based RL algorithms in \cite{Perrusquia2020,Radac2020,He2020} can be applied to control UAVs, provided that the UAV model satisfies the assumptions in \cite{Perrusquia2020,Radac2020,He2020}.

	DRL adopts both deep learning (DL) and reinforcement learning (RL) principles, where deep neural networks in DL are used to approximate Q-functions (or value functions) in RL and the RL agent learns the optimal strategy that maximizes the long-term cumulative rewards.
	 Most of the DRL algorithms can be classified into two categories: on-policy and off-policy methods. On-policy algorithms, such as trust region policy optimization (TRPO) \cite{Schulman2015} and proximal policy optimization (PPO) \cite{Schulman2017}, attempt to evaluate and improve the policy used to select actions. Off-policy algorithms, such as the deep Q-network (DQN) \cite{Mnih2013} and deep deterministic policy gradient (DDPG) \cite{Lillicrap2016}, evaluate and then improve policies different from those used to generate the data. 
	 It is known that off-policy methods are more sample-efficient compared to on-policy methods, since in off-policy methods, the past experiences performed by any policy can be used for learning \cite{Seijen2009, Sutton2018a_Book}.
	
	Soft actor-critic (SAC) is a class of off-policy DRL algorithms, which optimizes the stochastic policy based on the maximum entropy framework \cite{Haarnoja2018a}. SAC can be applied to various environments to achieve the state-of-the-art performance with respect to large continuous state and action spaces. 
%
SAC has advantages in terms of exploration and robustness compared with other DRL algorithms. That is, SAC outperforms earlier off-policy and on-policy DRL methods such as DQN and DDPG in terms of learning speed and cumulative reward \cite{Haarnoja2018a}. Note that SAC and other off-policy DRL algorithms include a technique called \emph{experience replay} to take advantage of past accumulated experiences \cite{Lin1993}. Although SAC with the experience replay can learn various environments with the advantages of exploration and robustness, the maximum entropy framework in SAC may degrade the optimality of learning outcomes after reaching the steady-state phase.

	Recently, hindsight experience replay (HER) was proposed in \cite{Andrychowicz2017} to improve the learning performance of DDPG. Specifically, HER is a sample-efficient replay method that enhances the performance of off-policy DRL algorithms by allowing the agent to learn from both failures and successes, similar to humans. Using the concept of \emph{goal}, HER provides the supplementary reward to the agent, which improves the optimality of the learning outcomes even if the goal is not achieved.	
	For HER in DDPG, the unshaped (binary or sparse) reward was used. However, this kind of reward is less informative for DRL agents to learn in environment with large continuous states and action spaces, which may decrease efficiency and speed of the learning process. Moreover, although DDPG is able to deal with environments with continuous state and action spaces, it suffers from instability, i.e., it may converge to unstable solutions or diverge, due to the high sensitivity to hyperparameters in DDPG  \cite{Matheron2019}. 
	

	In this paper, we propose a class of deep reinforcement learning (DRL) algorithms, SACHER, i.e., soft actor-critic (SAC) with hindsight experience replay (HER). As mentioned above, SAC outperforms earlier DRL algorithms including DQN and DDPG in terms of exploration, robustness, and learning performance. However, in SAC, maximizing the entropy-augmented objective function may degrade the optimality of the learning outcomes. We resolve this limitation by proposing SACHER, which improves the learning performance of SAC via HER. More precisely, SACHER achieves the desired optimal outcomes faster and more accurately than SAC, since HER improves the sample efficiency of SAC. Also, SACHER is able to avoid instability in DDPG with HER.

The main distinction between SACHER and SAC is that unlike SAC, SACHER stores a transition tuple of current and next states, action, reward, and initial goal. Next, by HER, SACHER samples additional goals from the states visited in the current episode. Although the transition tuple with the initial goal has a poor reward, SACHER obtains a supplementary reward for each additional goal and then stores the transition tuple of current and next states, the action, the supplementary reward, and the additional goal. By iterating this process, SACHER is able to generate learning outcomes more accurately and faster than SAC.

	We apply SACHER to the navigation and control problem of unmanned aerial vehicles (UAVs), where SACHER generates the optimal navigation path for the UAV under various obstacles. In simulation benchmark results, we demonstrate the effectiveness of SACHER in terms of the collision avoidance performance and the cumulative reward in UAV operation by comparing them with state-of-the-art DRL algorithms, SAC and DDPG. Note that SACHER in UAV navigation and control problems can be applied to arbitrary models of UAVs, since SACHER does not require specific information of UAV models and types of controllers.
	
	In summary, the main contributions of the paper can be stated as follows:
	\begin{enumerate}[(a)]
	  \item  We apply HER to SAC and propose SACHER to improve the learning performance of SAC;
	\item We apply SACHER to the navigation and control problem of UAVs under various obstacles. 
	\end{enumerate}
	We mention that SACHER cannot be viewed as a trivial application of HER to SAC. Specifically, the main technical challenges of SACHER and its application to the UAV navigation and control problems are as follows:
	\begin{enumerate}[(i)]
		\item The goal of HER causes intricacies throughout the entire SAC structure; 
		\item The unshaped (binary or sparse) reward of HER in \cite{Andrychowicz2017} is not directly applicable to environments with large continuous states and action spaces;
		\item Various model-based approaches in UAV navigation and control problems may not be able to deal with complex obstacles and operation constraints.
	\end{enumerate} 
	Regarding (i), the overall structure of SAC should be modified to consider the goal of HER. Unlike SAC, the SACHER agent has to select an action by considering not only the state but also the goal. This leads to intricacies while evaluating from policy to Q-function for implementing of SACHER. As for (ii), HER only returns binary or sparse rewards to the SACHER agent, which may not be applicable in environment with large continuous states and action spaces. We address (i) by merging the goal into the state space of SAC. In addition, (ii) is addressed by combining the unshaped reward with auxiliary reward, where the auxiliary reward is shaped as a quadratic function of state and goal. In (iii), the model-based approaches in UAV control and navigation problems are dependent on specific types of obstacles, UAVs, and objective functions. Moreover, it is necessary to compute associated gradients and/or complex Hamilton-Jacobi PDEs to obtain the optimal navigation path under obstacles \cite{Sundar1997,Altarovici2013,Fisac2015,Wang2020}. In this paper, we apply SACHER to address (iii). Specifically, by designing an appropriate UAV environment, SACHER is able to generate the optimal navigation path for the UAV to avoid collisions and obstacles regardless of UAV models and/or controllers. 
		
		The paper is organized as follows. SACHER is proposed in Section \ref{section2}. The simulation setup and environment design of the SACHER-based UAV navigation and control problem is discussed in Section \ref{section3}. The simulation results are provided in Section \ref{section4}. We conclude this paper in Section \ref{section5}.

	\section{Soft Actor-Critic Algorithm with Hindsight Experience Replay}\label{section2} 
	
	In this section, we first describe SAC and HER studied in \cite{Haarnoja2018a} and \cite{Andrychowicz2017}, respectively. Then we propose SACHER and explain its detailed algorithm and implementation.

	\subsection{Soft Actor-Critic (SAC) Algorithm}
	\label{section2.A}
	As shown in \cite{Haarnoja2018a}, SAC is a class of the maximum entropy DRL algorithms, which optimizes the following objective function:
	\begin{align}\label{eq1}
	J(\pi) = \sum_{t=0}^T{\mathbb{E}}_{(s_t,a_t)\sim \rho_\pi} \left[ r(s_t, a_t) + \alpha {\mathcal{H}}(\pi(\cdot|s_t))\right],
	\end{align}
	where $r_t=r(s_t, a_t)$ is the reward obtained when the SACHER agent executes the action $a_t \in \mathcal{A}$ in the state $s_t \in \mathcal{S}$, $\pi$ is the policy, $\rho_\pi$ is the joint distribution over states and actions induced by the policy $\pi$, $\alpha$ is the temperature weight of the entropy term $\mathcal{H}$, and ${\mathcal{H}}(\pi(\cdot|s_t)) = -\mathbb{E}_{\pi}[\log \pi(\cdot|s_t)] = - \int_{\mathcal{A}} \pi(a|s_t) \log \pi(a|s_t) da$ is the entropy. Here, $\mathcal{A}$ and $\mathcal{S}$ denote action and state spaces, respectively.

	The main objective of SAC is to find the optimal policy $\pi^*$ that maximizes the entropy-augmented reward function $J(\pi)$ in \eqref{eq1}, which requires the soft policy iteration of soft Q- and value functions. The soft Q-function satisfies the following soft Bellman equation:
	\begin{align}\label{12334}
	Q(s_t,a_t) := r(s_t, a_t) + \gamma{\mathbb{E}}_{s_{t+1}\sim p}\left[V({s_{t+1}})\right],
	\end{align}
	where
	\begin{align}\label{eq_vf}
	V(s_t) := {\mathbb{E}}_{a_{t}\sim \pi}\left[Q(s_t,a_t)-\alpha\log\pi(a_t|s_t)\right]
	\end{align}
	is the soft value function, and $p= p(s_{t+1}|a_t, s_t)$ is the state transition probability, which represents the probability density of the next state $s_{t+1} \in \mathcal{S}$ given the current state $s_{t} \in \mathcal{S}$ and the action $a_{t} \in \mathcal{A}$. We can evaluate the soft Q value of a fixed policy $\pi$ by applying the Bellman equation in \eqref{12334} to each time step, which is the so-called soft policy evaluation in the soft policy iteration. The objective function for the soft policy iteration can be written as follows:
	\begin{align}\label{eq_111223}
	J_\pi(\pi) 
	&= {\mathbb{E}}_{s_{0}\sim p}\left[D_{KL}\left(\pi(\cdot|s_0)~\bigg|\bigg|~\exp\left(\frac{1}{\alpha}Q(s_0,\cdot)\right)\right)\right],
	\end{align}
	where $D_{KL}$ denotes the Kullback-Leibler (KL) divergence. We can easily show that the original maximization problem of \eqref{eq1} is equivalent to the minimization of \eqref{eq_111223} due to the definition of $D_{KL}$ \cite{Haarnoja2018a}. The minimization of the objective $J_\pi$ in \eqref{eq_111223} with respect to the policy $\pi$ is called as soft policy improvement in the soft policy iteration.  Note that applying directly the above soft policy iteration to large continuous and action spaces requires a certain type of practical approximations \cite{Haarnoja2018a}.

	Instead of executing the policy iteration until convergence, parameterized neural networks for the Q-function and the policy are used as function approximators. The soft Q-network is parameterized by $\theta$, where the parameter for the soft policy network is denoted by $\phi$. Then the soft Q-function parameters $\theta$ can be optimized by minimizing the squared soft Bellman residual given by 
	\begin{align}\label{loss1}
	J_Q(\theta) &= {\mathbb{E}}_{(s_t,a_t)\sim \mathcal{D}} \left[\frac{1}{2}(Q_\theta(s_t,a_t)-(r(s_t,a_t)\right.\\&~~~~+\gamma  {\mathbb{E}}_{s_{t+1}}V_{\bar{\theta}}(s_{t+1}) ))^2 \bigg], \nonumber
	\end{align}
	where {$\mathcal{D}$} denotes the replay buffer, and $\bar{\theta}$ is the target Q-function parameter. It should be noted that the soft value function is also parameterized by the soft Q-function parameter $\theta$, due to the relation with the Q-function in \eqref{eq_vf}. The policy network, parameterized by $\phi$, can be learned by minimizing the expected KL divergence in \eqref{eq_111223}:
	\begin{align}\label{loss2}
	J_\pi (\phi) =  {\mathbb{E}}_{s_t\sim \mathcal{D}} \left[ {\mathbb{E}}_{a_t\sim\pi_\phi} \left[\alpha\log\pi_\phi(a_t|s_t)-Q_\theta(s_t,a_t)\right]\right].
	\end{align}

	Finally, one can minimize the loss functions in \eqref{loss1} and \eqref{loss2} by using the stochastic gradient descent (SGD) method. For SGD, we use two soft Q-networks parameterized by $\theta_i$, $i=1,2$, which are trained independently to optimize the soft bellman residual in \eqref{loss1}. The minimum of the two soft Q-functions is used for SGD to minimize the loss functions in \eqref{loss1} and \eqref{loss2}. As mentioned in Section \ref{section1}, since SAC is an off-policy maximum entropy-based algorithm, it has advantages in terms of exploration and robustness. The outstanding performance of SAC is demonstrated in  \cite{Haarnoja2018a}. The results of SAC outperform those of the earlier off-policy and on-policy DRL methods (including both DDPG and PPO) in terms of learning speed and cumulative reward. 
	
	\subsection{Hindsight Experience Replay (HER)}
	The main idea of hindsight experience replay (HER) in \cite{Andrychowicz2017} is to allow the DRL agents to learn from both failures and successes, similar to humans. To achieve this, HER employs the concept of a \emph{goal} $g \in \mathcal{G}$ used in \cite{Schaul2015}, where $g$ represents the goal (or objective) that the DRL agent has to achieve in the environment and $\mathcal{G}$ represents the corresponding goal space. Then the modified reward function $r_t = r(s_t, a_t, g)$ is defined as a function of not only the state and action, but also the goal. The closer the state $s_t$ is to the goal $g$, the greater the reward the DRL agent receives. 
		
		The detailed process of HER is as follows. After executing the environment steps in each episode, HER has the knowledge of the visited states $\zeta=\{s_0, s_1, \ldots,s_T\}$. Based on the knowledge, HER first stores in the replay buffer $\mathcal{D}$ every transition tuple $(s_t, a_t, r_t, s_{t+1})$ together with the original goal $g$. Then, HER stores extra transition tuples $(s_t, a_t, r'_t, s_{t+1})$ together with $g'\in \varphi$, where $\varphi=\{g'_1, g'_2, \ldots, g'_m\}$ is a set of the additional goal uniformly sampled from the visited states $\zeta$ in the current episode. From this process, HER gives some supplementary rewards $r'_t = r(s_t, a_t, g')$ to the DRL agent, although the original goal $g$ is not achieved in that episode and the SACHER agent may get the poor reward. This process enhances the original DRL algorithm in terms of the learning speed and success rate of reaching the goal.

	\begin{algorithm*}[t]
		\caption{Soft actor-critic with hindsight experience replay (SACHER)}
		\label{SACHER}
		\begin{algorithmic}[1]
			\State{Initialize Q-function parameters $\theta_1$, $\theta_2$, policy parameters $\phi$, empty replay buffer {$\mathcal{D}\leftarrow \emptyset$}}
			\State{Initialize target Q-function parameters $\bar{\theta}_1\leftarrow \theta_1$,$\bar{\theta}_2\leftarrow \theta_2$} 
			\State{Initialize goal $g$ from the goal space $\mathcal{G}$}
			\For {each episode}
			\State{Sample initial state $s_0$}
			\For {each environment step}
			\State{Select action based on the current policy $a_t \sim \pi_\phi(a_t|s_t,g)$}
			\State{Execute action $a_t$ and observe new state $s_{t+1} \sim p(s_{t+1}|a_t,s_t, g)$} 
			\EndFor
			\For {each environment step}
			\State{Obtain reward $r_t=r(s_t,a_t,g)$ from the environment}
			\State{Store transition tuple $(s_t,a_t,r_t,s_{t+1},g)$ in the replay buffer $\mathcal{D}$} \Comment{Standard experience replay}
			\State{Sample additional goals $\varphi=\{g'_1, g'_2, \ldots, g'_m\}$ from the states $\zeta=\{s_0, s_1, \ldots,s_T\}$ } 
			\For{each additional goal $g' \in \varphi$}
			\State{Obtain supplementary reward $r'_t=r(s_t,a_t,g')$ from the environment}
			\State{Store transition tuple $(s_t,a_t,r'_t,s_{t+1},g')$ in the replay buffer $\mathcal{D}$ }\Comment{Hindsight experience replay}
			\EndFor
			\EndFor
			\State{Sample a minibatch from the replay buffer $\mathcal{D}$} 
			\For{each gradient step}
			\State{Update the Q-function parameters $\theta_i \leftarrow \theta_i-\lambda_Q\hat{\nabla}_{\theta_i}J_Q(\theta_i)~\text{for~}i\in\{1, 2\} $}
			\vspace{1mm}
			\State{Update the soft policy parameters $\phi \leftarrow \phi-\lambda_\pi\hat{\nabla}_{\phi}J_\pi(\phi)$}
			\vspace{1mm}
			\State{Adjust temperature weight $\alpha\leftarrow \alpha-\lambda\hat{\nabla}_{\alpha}J(\alpha)$}
			\vspace{1mm}
			\State{Update the target Q-function parameters $\bar{\theta}_i \leftarrow \tau\bar{\theta}_i+(1-\tau)\bar{\theta}_i~\text{for~}i\in\{1, 2\} $}	
			\EndFor
			\EndFor
		\end{algorithmic}
	\end{algorithm*}
	
	\subsection{SACHER: Soft Actor-Critic Algorithm with Hindsight Experience Replay}
		In this subsection, we propose SACHER in Algorithm \ref{SACHER} and address (i) in Section \ref{section1}. In detail, SACHER first empties the replay buffer $\mathcal{D}$, and then randomly initializes two soft Q-networks parameterized by $\theta_1$ and $\theta_2$ and the soft policy network parameterized by $\phi$. Moreover, the goal $g$ is initialized from the goal space $\mathcal{G}$. Then SACHER updates two target Q-networks parameterized by $\bar{\theta}_1$ and $\bar{\theta}_2$. For each episode, SACHER randomly chooses an initial state $s_0$. Then SACHER interacts with the environment during the environment steps by executing the actions based on the policy $\pi$ and observing the states $\zeta=\{s_0, s_1, \ldots,s_T\}$. For each environment step, SACHER obtains the reward $r_t=r(s_t,a_t,g)$, and stores the transition tuple $(s_t,a_t,r_t,s_{t+1},g)$ in the replay buffer $\mathcal{D}$, which corresponds to standard experience replay. Next, the additional goals $\varphi=\{g'_1, g'_2, \ldots, g'_m\}$ are  uniformly sampled from the set of visited states $\zeta=\{s_0, s_1, \ldots,s_T\}$ in the current episode. For each additional goal $g' \in \varphi$, SACHER obtains the supplementary reward $r'_t=r(s_t,a_t,g')$, and then stores the transition tuple $(s_t,a_t,r'_t,s_{t+1},g')$ in the replay buffer $\mathcal{D}$, which corresponds to HER. Finally, the soft Q-networks, the soft policy network, and the temperature weight $\alpha$ are optimized based on the SGD method. For each gradient step of SGD, the Q-function parameters $\theta_1$ and $\theta_2$, and the policy parameter $\phi$ are optimized to minimize the loss functions in \eqref{loss1} and \eqref{loss2}, respectively. The adjusted temperature weight $\alpha$ is optimized in each gradient step of SGD to minimize the following objective:
		\begin{align*}
		J(\alpha)= {\mathbb{E}}_{a_{t}\sim \pi}\left[-\alpha\log\pi(a_t|s_t)-\alpha\bar{\mathcal{H}}\right],
		\end{align*}
		where $\bar{\mathcal{H}}$ is desired target entropy. The target Q-function parameters $\bar{\theta}_1$ and $\bar{\theta}_2$ are updated by the exponentially moving average method with the smoothing constant $\tau$. After the iteration of the above learning process, SACHER gives the optimized soft Q-function parameters, the soft policy parameter $\phi$, and the corresponding transition tuples.

\begin{figure*}[t!]
	\begin{center}
		\includegraphics[width=17cm]{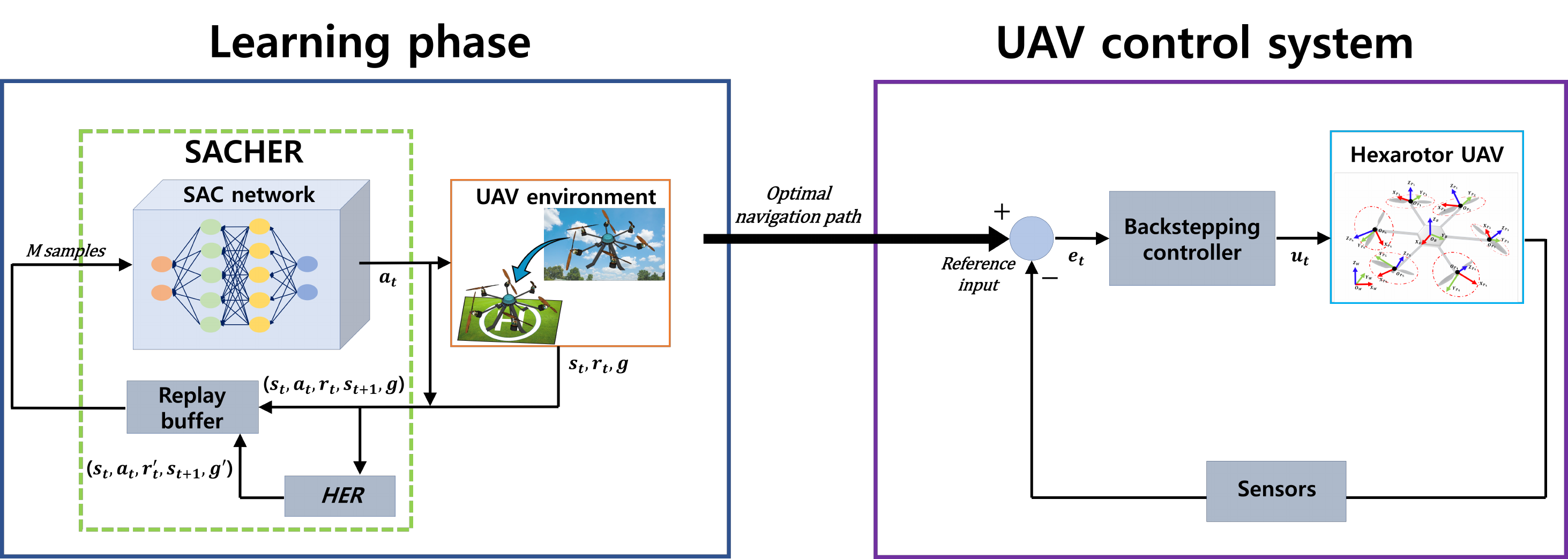}
		\caption{\label{fig_NN} Framework of SACHER-based UAV navigation and control. In the learning phase, SACHER learns the UAV environment, and then generates the optimal navigation path. After completing the learning phase, the output of SACHER is considered as the reference input for the UAV control system. }
	\end{center}
\end{figure*}

	The preceding analysis indicates that in SACHER, the concept of the goal space $\mathcal{G}$ and the additional storage of the transition tuple $(s_t,a_t,r'_t,s_{t+1},g')$ are introduced by HER. Therefore, as stated in (i) of Section \ref{section1}, the entire framework of SACHER becomes more complicated than the structure of SAC due to the implementation of goals in HER. Specifically, let $\mathcal{S}$ be the state space, $\mathcal{A}$ the action space, and $\mathcal{G}$ the goal space. Then the corresponding reward $r : \mathcal{S} \times \mathcal{A} \times \mathcal{G} \rightarrow \mathbb{R}$ is defined as a function of state, action, and goal. In every environment step, the SACHER agent selects the action based on the policy $\pi: \mathcal{S} \times \mathcal{G}$ $\rightarrow\mathcal{A}$ and then observes the next state $s_{t+1}$ based on the state transition probability $p(s_{t+1}|a_t,s_t,g)$. The Q-function now depends on the state-action pair together and the goal, i.e., $Q(s_t, a_t, g) = \mathbb{E}[R_t|s_t, a_t, g]$. These complex procedures cause intricacies while evaluating the policy iteration, the Q-function, as well as the stochastic gradient descent. We resolve this difficulty by merging the goal $g$ into the state $s_t$, since the goal is fixed and not changed during the entire learning process. 
	
		
We now discuss the convergence of SACHER. Note that HER in SACHER does not affect the network structure of SAC. Moreover, HER is related to the concept of the goal space $\mathcal{G}$ and the additional storage of the transition tuple $(s_t,a_t,r'_t,s_{t+1},g')$. One notable modification of SACHER from SAC is the structure of the Markov decision process (MDP) in SACHER. Specifically, by HER, the MDP of SACHER is represented by the 6-tuple $(\mathcal{S},\mathcal{A},\mathcal{R},p,\gamma, \mathcal{G})$, where the goal space $\mathcal{G}$ is augmented into the MDP of SAC represented by the $5$-tuple $(\mathcal{S},\mathcal{A},\mathcal{R},p,\gamma)$. Then given the goal space, the MDP of SACHER can be modified by the $5$-tuple $(\bar{\mathcal{S}},\mathcal{A},\mathcal{R},p,\gamma)$ with $\bar{\mathcal{S}} = \mathcal{S} \times \mathcal{G}$. This implies that the MDP of SACHER represented by the modified $5$-tuple $(\bar{\mathcal{S}},\mathcal{A},\mathcal{R},p,\gamma)$ can be viewed as the MDP of SAC represented by the $5$-tuple. We now apply the convergence analysis of SAC given in \cite[Theorem 1]{Haarnoja2018a} to SACHER. That is, we can easily modify \cite[Theorem 1]{Haarnoja2018a} to show the convergence of SACHER by representing the MDP of SACHER as the modified $5$-tuple $(\bar{\mathcal{S}},\mathcal{A},\mathcal{R},p,\gamma)$. This implies the convergence of SACHER.

	\section{Simulation Setup and Environment Design}\label{section3}
	
		In this section, we first state the detailed simulation setup for the SACHER-based navigation and control problem of UAVs. Then we design two environments; Environment I is for the case without obstacles and Environment II is for the case with obstacles. Note that these two environment design analysis addresses (ii) in Section \ref{section1}.
		
		\subsection{SACHER-based UAV Navigation and Control}

	The entire framework of the SACHER-based UAV navigation and control system is described in Fig. \ref{fig_NN}. In the learning phase, based on the UAV environment design, SACHER learns the corresponding environment and then generates the optimal navigation path for UAVs via Algorithm \ref{SACHER}. After completing the learning phase, the output of SACHER is considered as the reference input for the UAV control system. In the UAV control system, the  tracking controller controls the UAV to follow the optimal navigation path generated by SACHER (see Fig. \ref{fig_NN}).

		In our simulations, the tilted-hexarotor UAV model in \cite{Lee_Franklin_2019,Rajappa2015a} is used. Moreover, we design the standard backstepping controller (see \cite{Isidori2013}) for the hexarotor UAV to track the optimal navigation path generated by SACHER. Note that in the learning phase, since SACHER does not require any information of UAV models (e.g., quadrotor, aircraft, or ground vehicles) and types of controllers, any (optimal/nonoptimal) nonlinear (or linear) controllers (with appropriate design modifications) can be used in Fig. \ref{fig_NN} instead of the hexarotor UAV and the backstepping controller. We use the hexarotor UAV and the backstepping controller not to show their control performances but to demonstrate the performance of SACHER.
			
	\subsection{Environment I: UAV without Obstacles}\label{section3B}
	We first consider a simple UAV environment, where the UAV lands in a landing area with the shortest path without any obstacles. To design this environment, we use a simple position and angle update equation of the UAV. Specifically, let $h = [x, y, z]^\top$ be the position, $\psi$ and $\dot{\psi}$ be the yaw angle and yaw angular speed, respectively, and $\tau$ be the yaw torque of the UAV. Then the position and angle update equation of the UAV can be written as follows:
	\begin{equation}
	\label{eq_aa}
	\begin{aligned}
	x_{t+1} &= x_t + v_1 \cos(\psi_{t}) \Delta t,~ y_{t+1} = y_t + v_1 \sin(\psi_{t}) \Delta t \\
	z_{t+1} &= z_t - v_2\Delta t \\
	\psi_{t+1} &= \psi_{t} + \dot{\psi}_{t} \Delta t,~ \dot{\psi}_{t+1} = \dot{\psi}_{t} + \tau_t \Delta t, 
	\end{aligned}
	\end{equation}
	where $v_1$ and $v_2$ are the positive constant velocities, and $\Delta t$ is the sampling time. We note that \eqref{eq_aa} does not depend on specific types of UAVs and environments.

	While interacting with environment, the SACHER agent observes the state $s=[x, y, z, \psi, \dot{\psi}]^\top$, and uses the yaw torque $\tau$ as an action ($a= \tau$). Note that the main objective of the UAV navigation and control problem is to reach the landing area on the xy-plane. We define $g=[g_x, g_y]^\top$ by the center of the landing area, which is the goal for Environments I and II. For simplicity, we assume that the landing area is the origin, i.e., $g = [0, 0]^\top$ (note that $g$ does not need to be the origin). Since the goal $g$ is defined on the xy-plane, whether the goal is achieved or not depends on states $x$ and $y$ updated by (\ref{eq_aa}). The landing area for the UAV environment is defined by a square located at the goal, i.e., 
	\begin{align}
	\label{eq_SSS}
	{\mathcal{L}}=\{h\in {\mathbb{R}}^3~|~ |x-g_x| \leq l_x, |y-g_y| \leq l_y, z = 0\},	
	\end{align}
	where $l_z$ and $l_y$ are the boundary constants. Then the reward function for each time step is defined by
	\begin{align}
	\label{eq_111}
	r(s_t,g) = r_1(s_t,g) + r_2(s_t, g),
	\end{align}
	where
	\begin{align}
	&r_1 = -k_1  \left((x_t-g_x)^2 +(y_t-g_y)^2 \right)-k_2(z_t)^2 \label{eq_quad}
	\\
	\label{eq_r2}
	&r_2 = \begin{cases}
	c_1, &\text{if $|x_t - g_x| \leq l_x, |y_t-g_y| \leq l_y$}, \\
	0, &\text{otherwise}
	\end{cases} 
	\end{align}
	with positive constant weights $k_1$ and $k_2$, and constant reward $c_1$. Here, \eqref{eq_quad} is the auxiliary reward, whereas \eqref{eq_r2} is the binary reward, which are used to address (ii) in Section \ref{section1}.
	
	The termination conditions of the UAV environment are set when (i) $z$ in \eqref{eq_aa} becomes zero, or (ii) $x$ and $y$ in \eqref{eq_aa} go into the landing area ${\mathcal{L}}$ of \eqref{eq_SSS}. The update model in \eqref{eq_aa} continues to descend with the constant velocity $v_2$ along with $z$-axis, regardless of the action of the SACHER agent. Therefore, under the environment of \eqref{eq_aa}, \eqref{eq_111}, and the above termination condition, the SACHER agent seeks to find the optimal path of the UAV reaching the landing area by maximizing the cumulative reward $R=\sum_{t=0}^{T}r_t$. Note that the cumulative reward $R$ is maximized when the second termination condition is satisfied with the fastest terminal time $T$, which provides the optimal navigation path for the UAV landing operation.

	As stated in (ii) of Section \ref{section1}, the approach used in HER only uses the binary reward as $r_2$ in \eqref{eq_r2}, which is not appropriate in environment with large continuous state and action spaces. With \eqref{eq_r2}, SACHER could not obtain any information before reaching the goal. In this case, even if the supplementary reward in HER exists, it is difficult for the agent to know the proper learning direction needed to achieve the goal. Therefore, we use the composite reward $r$ in \eqref{eq_111}, which includes the auxiliary reward in \eqref{eq_quad}, where \eqref{eq_quad} is shaped as a quadratic function of state and goal. In \eqref{eq_r2}, although the binary reward returns the zero reward when the goal is not achieved, the auxiliary reward provides another reward as a guideline of SACHER, indicating the degree of goal achievement. By using the composite reward in \eqref{eq_111}, we can prevent SACHER from learning through the wrong direction and reduce the number of episodes required for learning.
		
	\begin{remark}\label{Remark_1}
	In the learning phase of Environment I, to generate the optimal navigation path by SACHER, we apply the following learning processes based on Algorithm \ref{SACHER}:
	\begin{enumerate}
	\setlength{\itemindent}{0.05in}
	\item[(s.1)] With $g=[g_x, g_y]^\top$, the SACHER agent observes the state $s = [x, y, z, \psi, \dot{\psi}]^\top$ from Environment I;
	\item[(s.2)] Given the action $a =\tau$ of SACHER, the next state $s_{t+1}$ is updated based on \eqref{eq_aa};
	\item[(s.3)] The SACHER agent receives the reward $r_t$ in \eqref{eq_111} from Environment I ((s.1)-(s.3) show the collection of the state transition tuple $(s_t,a_t,r_t,s_{t+1},g)$);
	\item[(s.4)] After the storage of the state transition tuple $(s_t,a_t,r_t,s_{t+1},g)$ in the replay buffer $\mathcal{D}$, SACHER uniformly samples $m$ additional goals $\varphi= \{g_i' = (g'_{x,i},g'_{y,i}),~i=1,\ldots,m\}$ from the visited $x$ and $y$ states $\{(x_t,y_t),~ t=0,\ldots,T\}$\footnote{In the implementation of HER in (s.4) and (s.5), we set $m=4$ for the number of the sampled additional goals (see Table \ref{Table2}).}$^{,}$\footnote{As mentioned above (\ref{eq_SSS}), since the goal $g$ is defined on the xy-plane, the goal achievement depends on states $x$ and $y$.};
	\item[(s.5)] SACHER obtains the supplementary regard $r'_t$ for each additional goal $g' \in \varphi$ from Environment I (using \eqref{eq_111}), and stores the transition tuple $(s_t,a_t,r'_t,s_{t+1},g')$ in the replay buffer $\mathcal{D}$;
	\item[(s.6)] Under terminal conditions (i) and (ii), the SACHER agent obtains the optimal policy that maximizes the cumulative reward $R=\sum_{t=0}^T r_t$ based on experiences stored in the replay buffer  $\mathcal{D}$.
	\end{enumerate}
	Note that (s.1)-(s.6) are repeated for sufficient large learning (training) episodes.\footnote{The number of learning episodes is dependent on the problem setup.} After completing the learning phase, the output of SACHER is the reference input of the SACHER-based UAV navigation and control system (see Fig. \ref{fig_NN}). $\hfill{\square}$
	\end{remark}

	\subsection{Environment II: UAV with Obstacles}
	We consider a more complex UAV operation, where the UAV tries to avoid obstacles while tracking the optimal navigation path reaching the landing area generated by SACHER.
	
	A cylindrical obstacle ${\mathcal{O}}_i, 1 \leq i \leq N,$ is represented by the following zero-sublevel set:
	\begin{align}\label{eq_ob}
	\mathcal{O}_i&:=\{h \in \mathbb{R}^3~|~(x-x_{o,i})^2+(y-y_{o,i})^2-r_{o,i}^2\leq 0\},
	\end{align}
	where $r_{o,i}$ is the radius of the obstacle, and $(x_{o,i}, y_{o,i})$ is its location  on the xy-plane. The reward for the combined UAV operation of landing and obstacle avoidance is defined by
	\begin{align}\label{eq123}
	\bar{r}(s_t,g) = r(s_t,g) + \sum_{i=1}^N p_{i}(s_t),
	\end{align}
	where 
	\begin{align*}
	p_{i} =\begin{cases}
	-c_2, &\text{if $(x_t - x_{o,i})^2 + (y_t-y_{o,i})^2 -r_{o,i}^2 \leq c_3$}, \\
	0, &\text{otherwise}
	\end{cases}
	\end{align*}
	with the penalty constants $c_2$ and $c_3$. Note that in \eqref{eq123}, $r$ is given in \eqref{eq_111}, which can address (ii) stated in Section \ref{section1} as discussed in Section \ref{section3B}. The penalty constant $c_2$ decreases the cumulative reward $\bar{R} = \sum_{t=0}^T \bar{r}_t$ when the UAV collides with the obstacles in \eqref{eq_ob}, and $c_3$ acts as a margin that prevents the UAV from colliding with the obstacles. Under the environment of \eqref{eq_aa}, \eqref{eq_ob}, \eqref{eq123}, and the termination conditions of Environment I, the SACHER agent seeks to find the optimal path reaching the landing area without collisions by maximizing the cumulative reward $\bar{R}$. 
	
	We note that the implementation of SACHER for Environment II is identical with that of Environment I in  Remark \ref{Remark_1}. That is, the SACHER agent in Environment II follows the same learning phase as Environment I to generate the optimal navigation path under obstacles, where the reward in \eqref{eq123} and the cumulative reward $\bar{R} = \sum_{t=0}^T \bar{r}_t$ have to be used instead of \eqref{eq_111} and $R=\sum_{t=0}^T r_t$ in Remark \ref{Remark_1}.

	\begin{table}[t!] 
		\caption{Environment parameters} 
		\centering 
		\begin{tabular}{l||   l} 
			\hline\hline   
			Parameter & Value  \\	\hline   
			$v_1$: velocity on xy-plane & $2$ \\
			$v_2$: velocity for z-axis$~~~~~~~~~~~$ &  $0.5~~~~~~~~~~~~~~~~~~~~~~~~~$ \\ 
			$\Delta t$: sampling time & $0.1$ \\
			$\tau_t$: yaw angle torque & $[-0.5, 0.5]$  \\
			$l_x$, $l_y$: boundary constants & $0.2, 0.2$ \\
			$k_1$, $k_2$: weight constants & $10^{-3}, 10^{-4}$\\
			$c_1$: reward constant & $10$ \\
			$c_2, c_3$: penalty constants & $10$, $0.2$\\
			\hline
		\end{tabular}
		\vspace{2mm}
		\label{Table1}
	\end{table}

	\begin{table}[t!] 
		\caption{SACHER hyperparameters} 
		\centering 
		\begin{tabular}{l||   l} 
			\hline\hline   
			Hyperparameter & Value  \\	\hline   
			optimizer for SGD & Adam \cite{Kingma2015} \\
			learning rate for optimizer $\lambda_Q, \lambda_\pi, \lambda$ $~~~~~~~~~~~$ &  $3\times 10^{-4}~~~~~~~~~~~~$\\ 
			discount factor $\gamma$ & $0.99$\\
			number of hidden layers& $2$  \\
			number of hidden units & $256$ \\
			minibatch size & $256$ \\
			target smoothing coefficient $\tau$ & $0.005$ \\	
			activation function & ReLU \\			
			replay buffer capacity & $10^6$ \\
			number of additional goals (HER) $m$ & $4$ \\
			target entropy $\bar{\mathcal{H}}$ & $-\text{dim}(\mathcal{A})=-1$
			\\\hline
		\end{tabular}
		\label{Table2}
	\end{table}
	
	%
	
	\begin{table*}[t!]
	\caption{\label{R_Table_1} \color{blue} Learning time (h) for simulations}
	\centering
	\begin{tabular}{c||c|c|c|c|c|c}
		\hline
		& Ant-v2 & Walker2d-v2 & Hopper-v2 & HalfCheetah-v2 & Env. I & Env. II \\\hline\hline
		SAC   & 6.22h        & 4.81h & 5.08h               & 4.63h    & 1.85h         & 1.93h \\
		DDPG   & 7.54h       & 5.61h & 6.12h               & 5.59h    & 2.59h           & 2.80h \\
		SACHER   & 6.11h & 4.75h & 5.05h      & 4.53h     & 1.71h  &1.91h
		\\\hline
	\end{tabular}
\end{table*}

	\begin{figure}[t!]
		\begin{center}
		\centering
	\includegraphics[width=9cm]{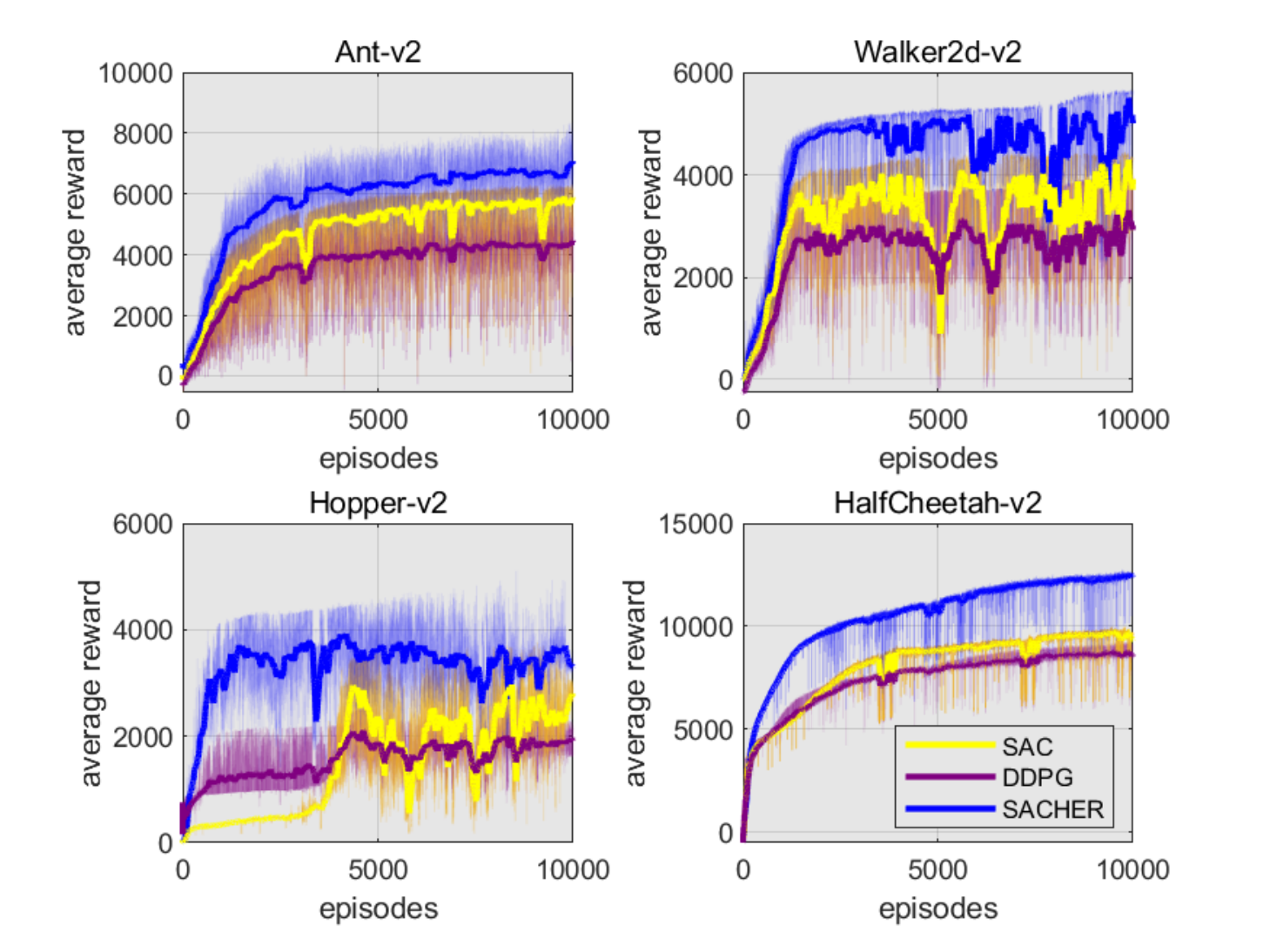}
	  \caption{\label{fig_mujoco_1} Benchmark results of SAC, DDPG and SACHER for the MuJoCo physics engine environment. The blue-colored line is the reward curve of SACHER, the yellow-colored line is the reward curve of SAC, and the purple-colored line is the reward curve of DDPG. In each plot, the solid curve represents the average reward, and the shaded area represents the minimum and maximum rewards in the four learning trials.}
		\end{center}
	\end{figure}

			\begin{figure*}[t!]
		\begin{center}
	\includegraphics[width=5.85cm]{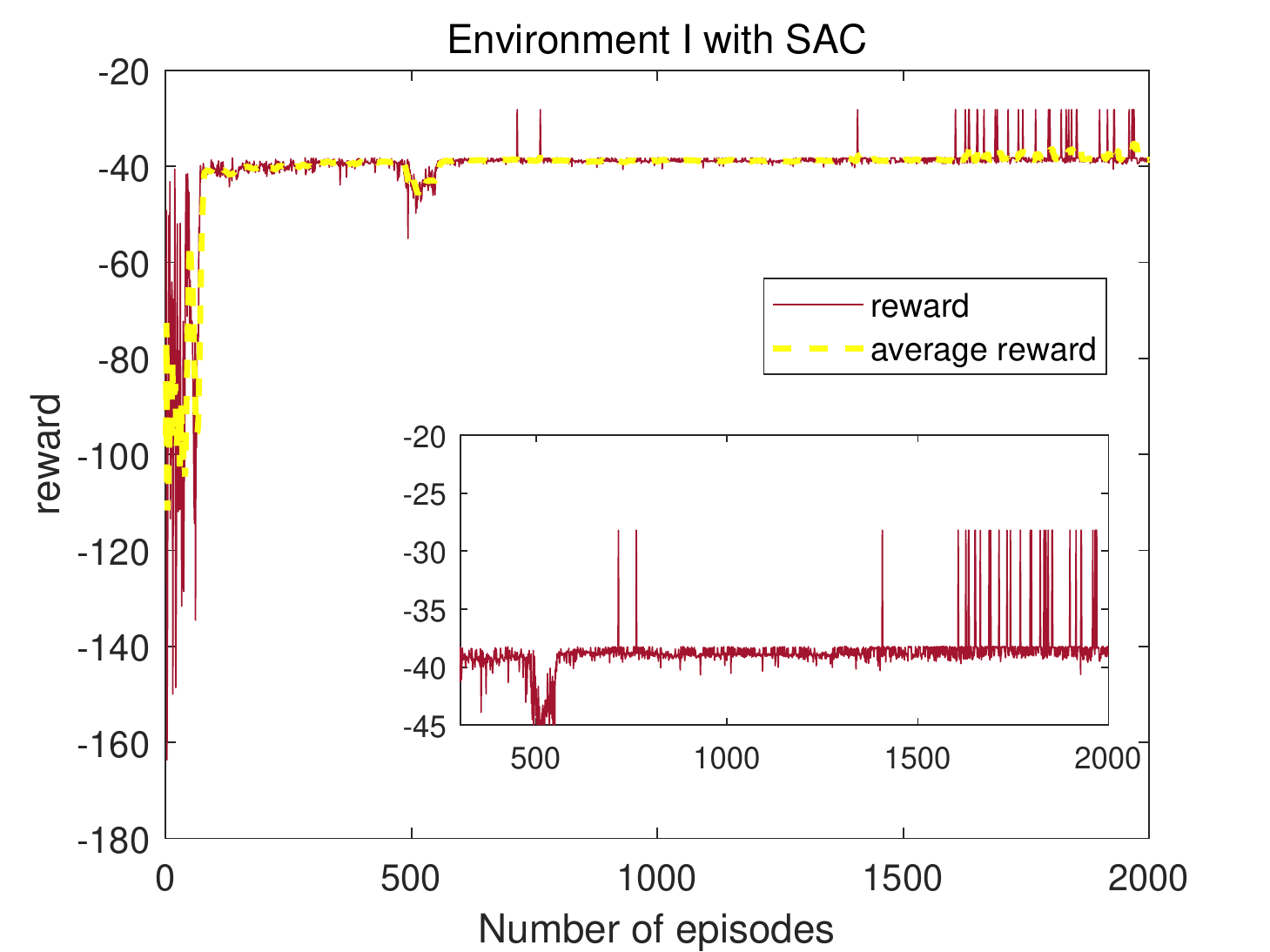} 
	\includegraphics[width=5.85cm]{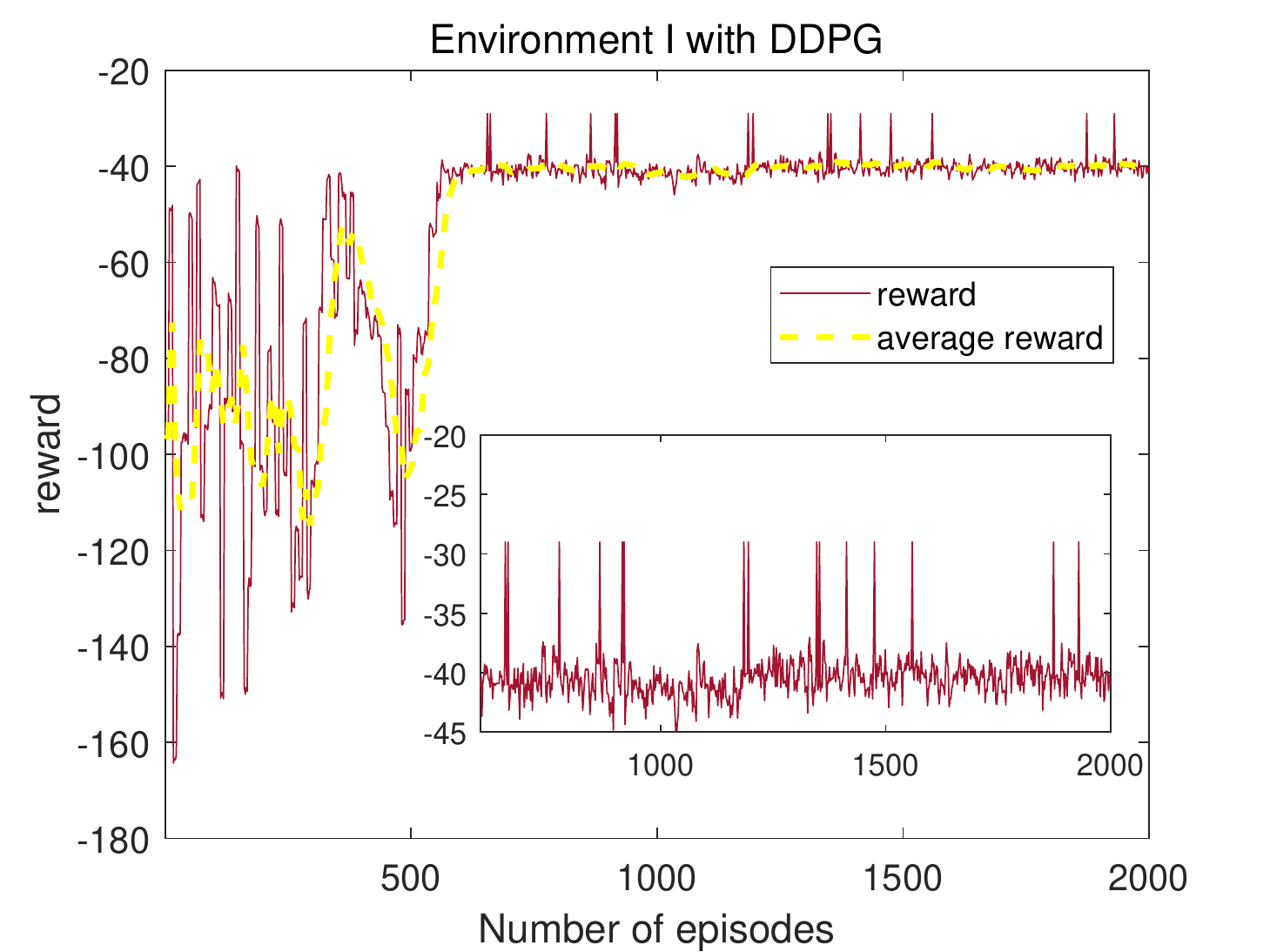}
	\includegraphics[width=5.85cm]{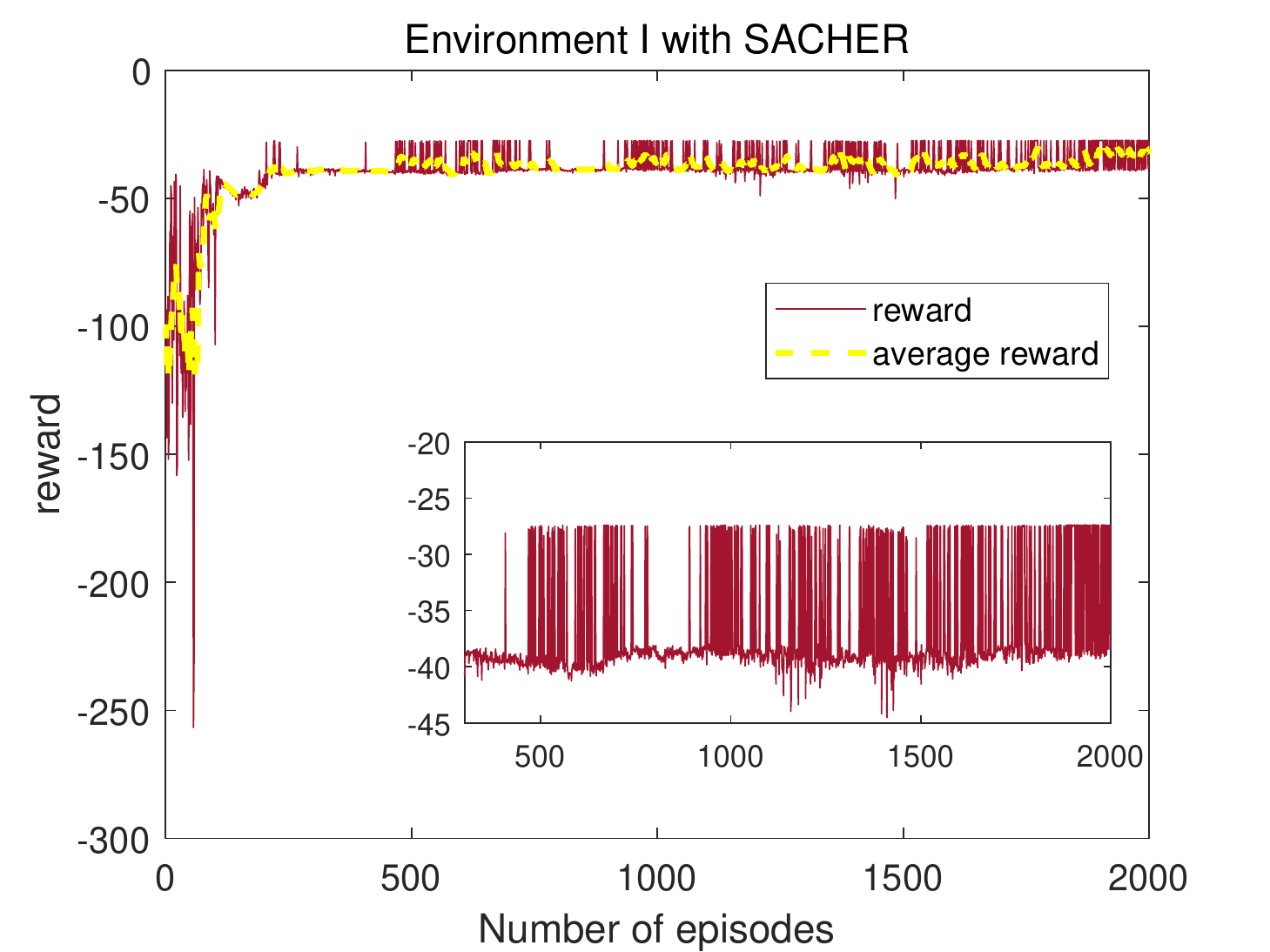}
			\caption{\label{fig1} Learning curves of SAC (left), DDPG (middle) and SACHER (right) in Environment I: The (brown-colored) solid line is the cumulative reward and the (yellow-colored) dotted line is the average of the last $20$ rewards. The subfigure illustrates the cumulative reward for each episode after steady-state. }
		\end{center}
	\end{figure*}

	\section{Simulation Results}\label{section4} 
	
	We provide the simulation results for navigation and control of the hexarotor UAV based on SACHER, where the detailed simulation setup is described in Section \ref{section3} and Fig. \ref{fig_NN}. For the comparison and validation of the learning performance of SACHER, we provide the simulation results of SAC \cite{Haarnoja2018a} and DDPG \cite{Lillicrap2016} under the same simulation environments.

	The simulations are performed via Python 3.6 on the Windows 10 operating system with Intel Core i9-10900X CPU and 64 GB RAM. For both Environments I and II, the initial state of the hexarotor UAV in the environment is fixed as $s_0 = [x_0, y_0, z_0, \psi_0, \dot{\psi}_0]^\top=[20, 20, 10, 5\pi/4, 0]^\top$ and the goal is fixed as $g=[g_x, g_y]^\top=[0, 0]$ (origin on xy-plane). 
	
	The values of the parameters in \eqref{eq_aa}-\eqref{eq_r2} and \eqref{eq123} are given in Table \ref{Table1}. The values of SACHER hyperparameters are provided in Table \ref{Table2}. In Environments I and II, HER selects $4$ additional goals ($m=4$) from the visited states in the current episode, and then stores the transition tuples with the corresponding supplementary rewards (see Remark \ref{Remark_1}). For Environment II, we consider the case when there exist $N=9$ obstacles, ${\mathcal{O}}_i, i=1, 2, \ldots, 9,$  where their locations $(x_{o,i}, y_{o,i})$ in \eqref{eq_ob} are $\{(5,5), (5,10), (5,15), (10,5), \ldots, (15, 10),(15, 15)\}$ with the same radius of $r_{o,i}=1$.

\begin{figure}[t!]
	\begin{center}
	\includegraphics[width=8.5cm]{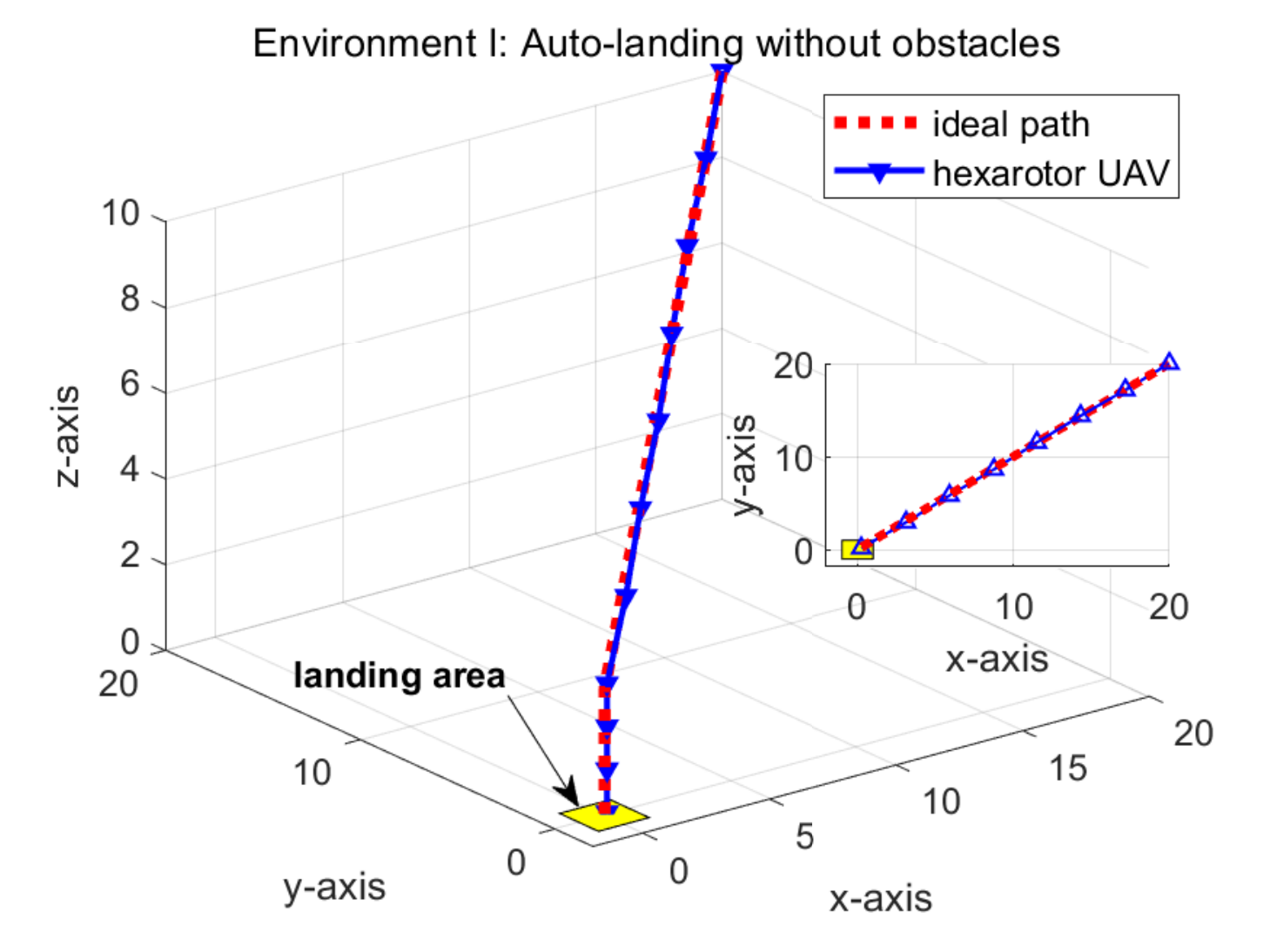}
	\caption{\label{fig2} The path of the hexarotor UAV with SACHER for Environment I. The path on the xy-plane is shown in the subfigure. The hexarotor UAV follows the optimal path generated by SACHER, which coincides with the ideal path without obstacles (obtained from the straightforward computation).}
		\end{center}
	\end{figure}

\begin{figure*}[t!]
		\begin{center}
	\includegraphics[width=5.85cm]{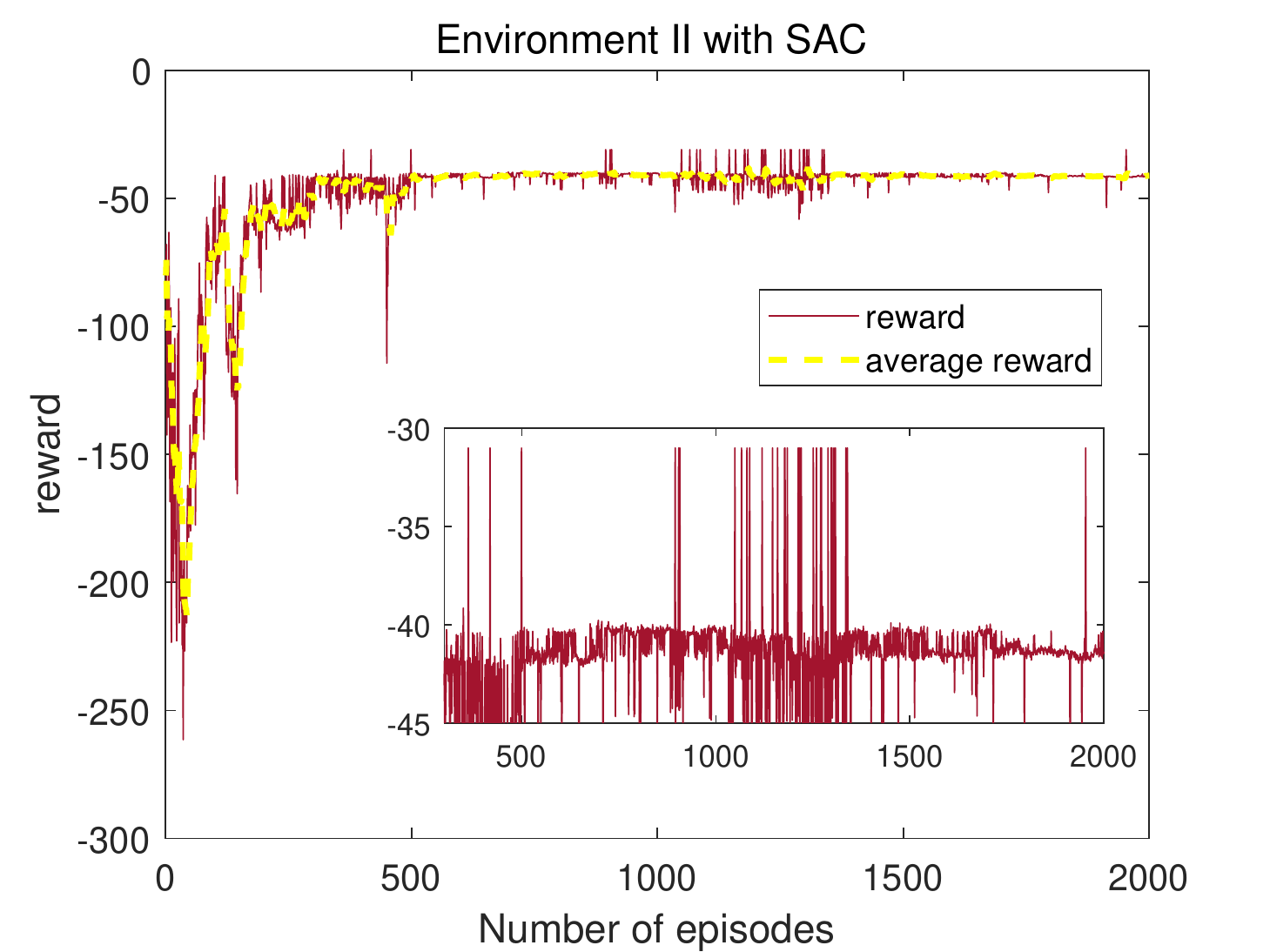}
	\includegraphics[width=5.85cm]{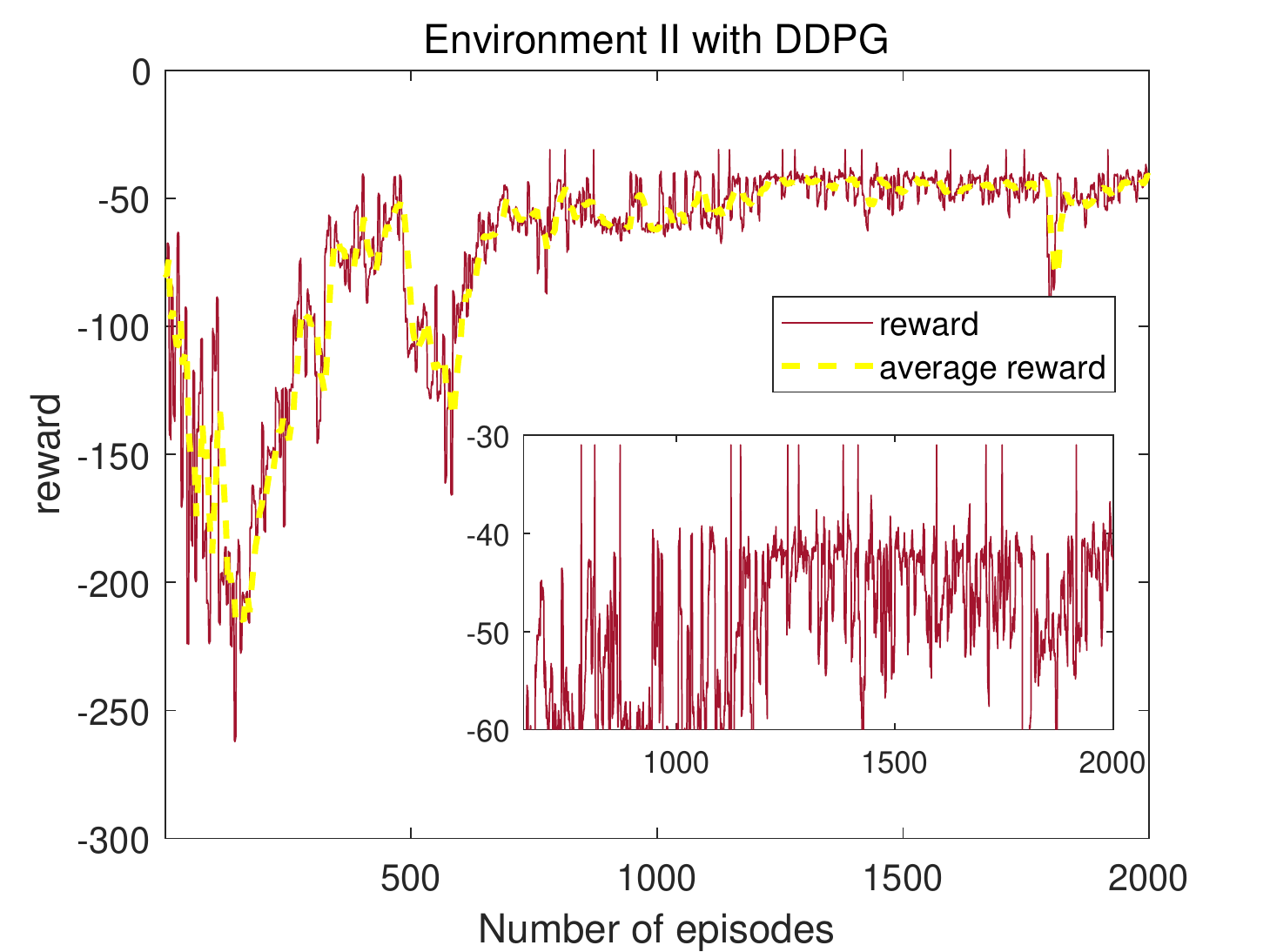}
	\includegraphics[width=5.85cm]{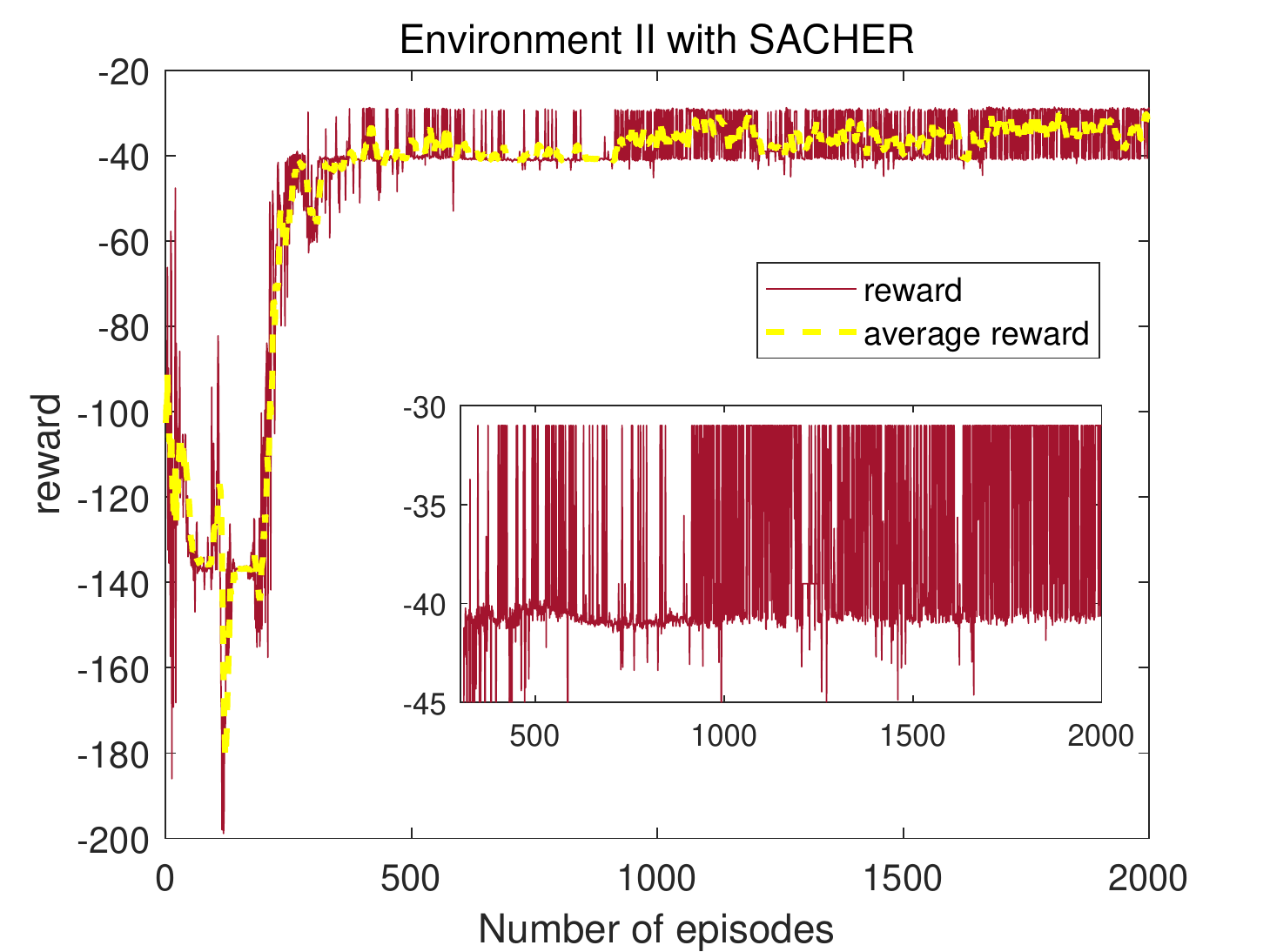}	
			\caption{\label{fig3} Learning curves of SAC (left), DDPG (middle) and SACHER (right) in Environment II. The (brown-colored) solid line is the cumulative reward and the (yellow-colored) dotted line is the average of the last $20$ rewards. The subfigure illustrates the cumulative reward for each episode after steady-state.}
		\end{center}
	\end{figure*}

\begin{figure}[t!]
		\begin{center}
		\includegraphics[width=9cm]{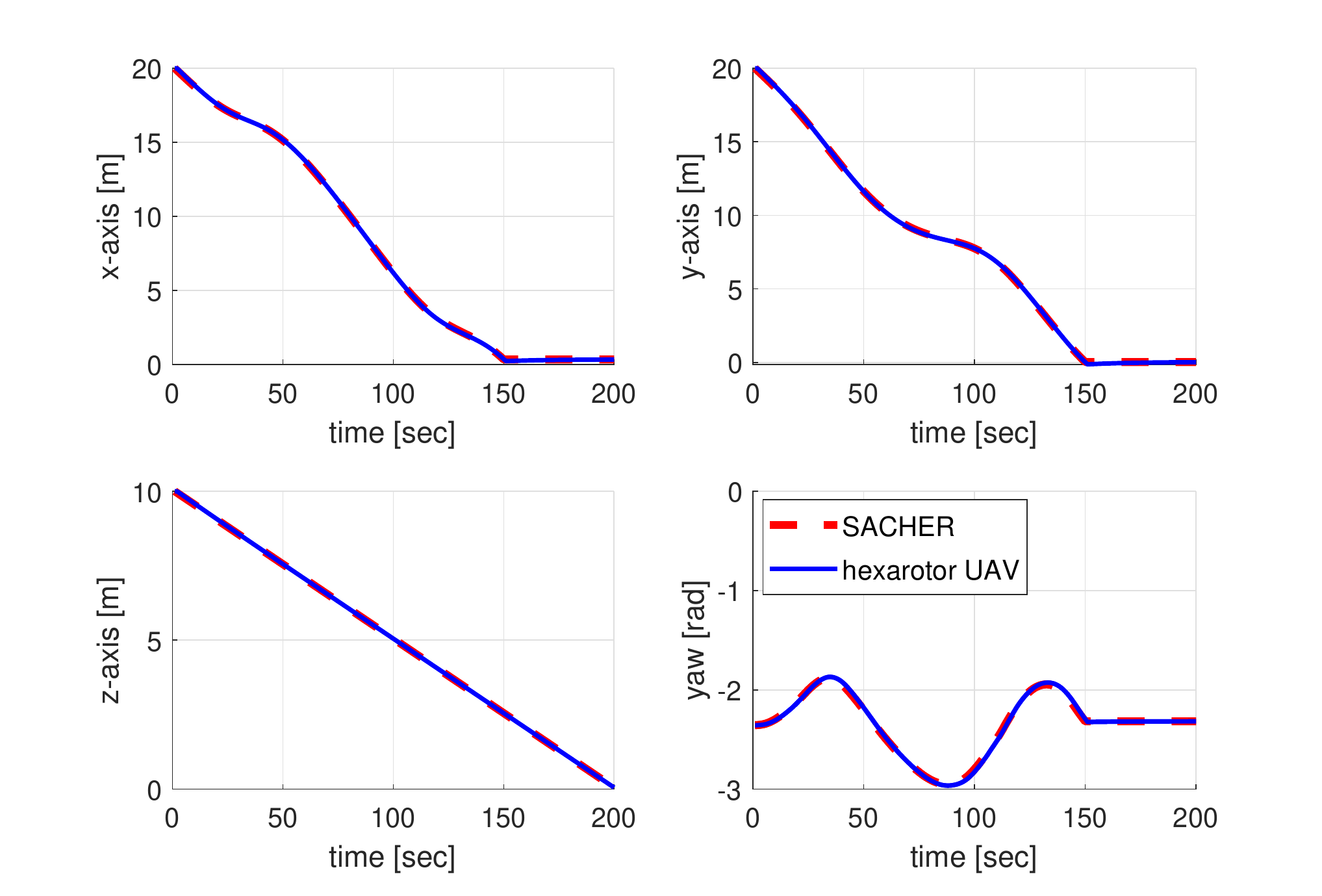}
		\includegraphics[width=9cm]{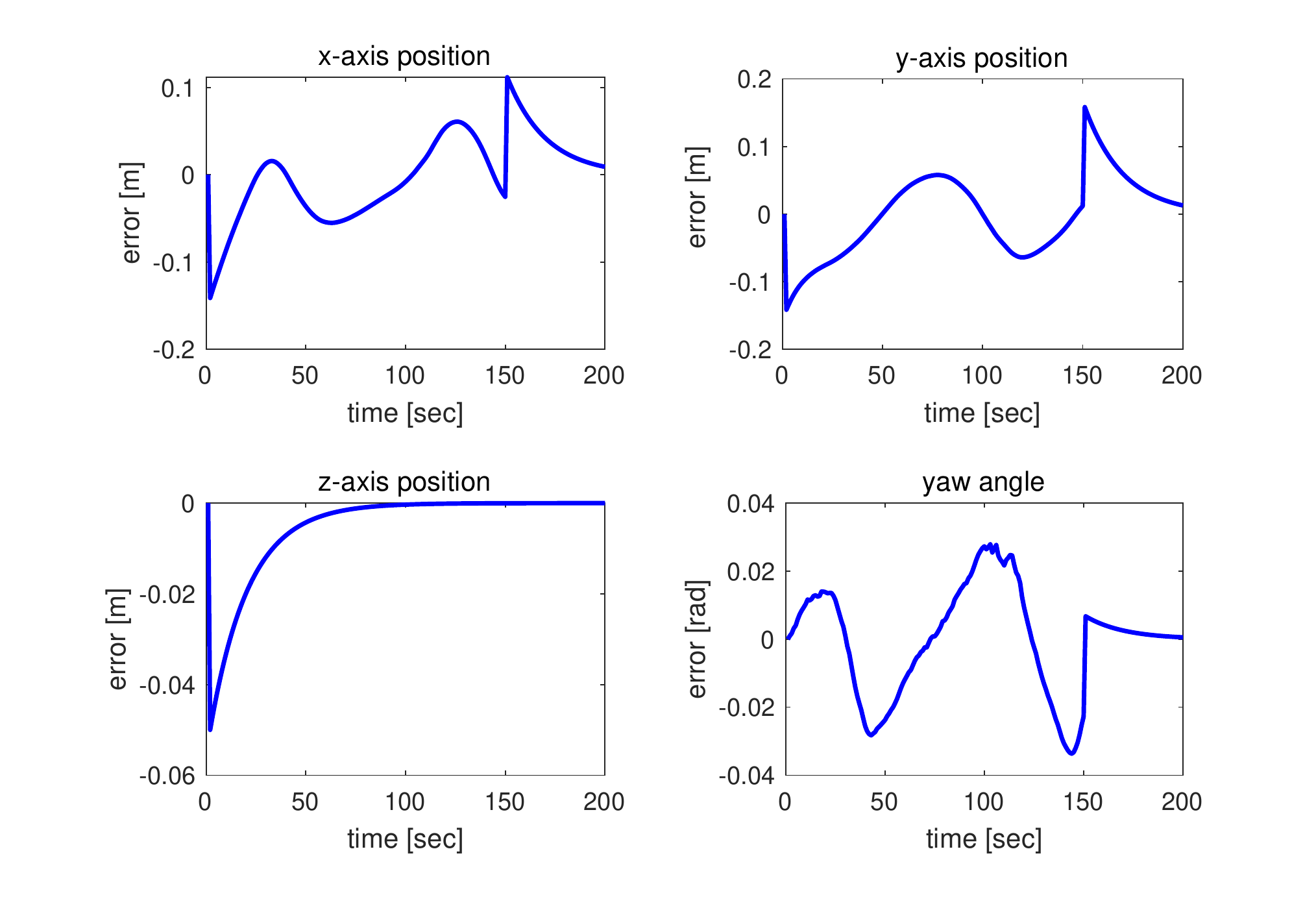}
			\caption{\label{fig7}The paths and tracking errors of the hexarotor UAV using the backstepping controller for Environment II. The (red-colored) dotted line is the optimal navigation path provided by SACHER, and the (blue-colored) solid line is the path of the hexarotor UAV controlled by the backstepping controller.}
		\end{center}
	\end{figure}

\begin{figure}[t!]
		\begin{center}
	\includegraphics[width=8.5cm]{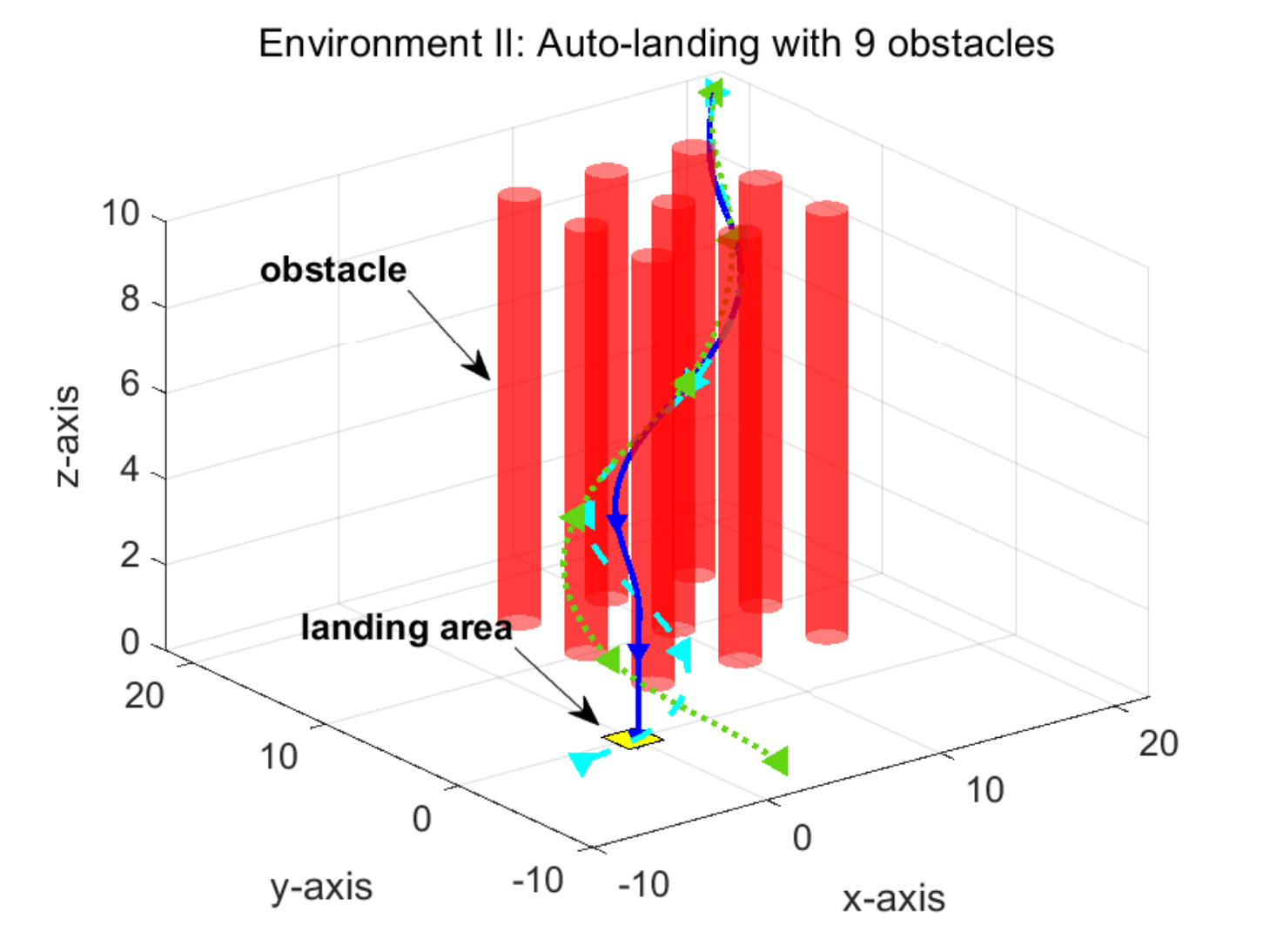}  
	  \includegraphics[width=8.5cm]{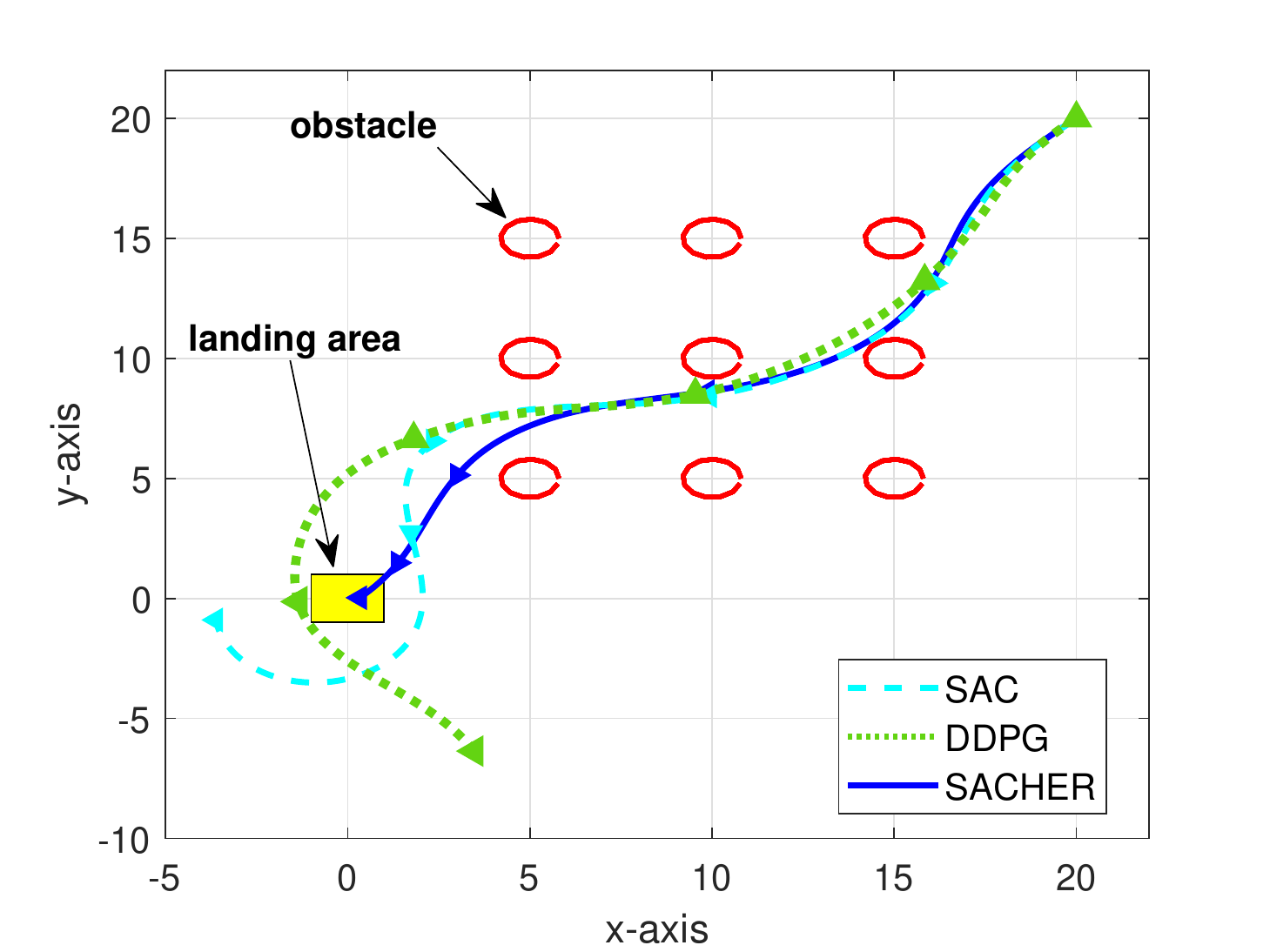}
			\caption{\label{fig5}{The paths of the hexarotor UAV with SAC, DDPG and SACHER for Environment II. The turquoise-colored dashed line is the path of the hexarotor UAV with SAC, the green-colored dotted line is the path of the hexarotor UAV with DDPG, and the blue-colored solid line is the path of the hexarotor UAV with SACHER. The result shows that the hexarotor UAV with SACHER successfully reaches the landing area, while the others do not.
			}}
		\end{center}
	\end{figure}

Table \ref{R_Table_1} provides the learning time of SAC \cite{Haarnoja2018a}, DDPG \cite{Lillicrap2016} and SACHER for MuJoCo physics engine (see Fig. \ref{fig_mujoco_1}) and Environments I-II (see Figs. \ref{fig1} and \ref{fig3}). From Table \ref{R_Table_1}, we can see that SAC and SACHER show faster learning time than that of DDPG. In Table \ref{R_Table_1}, although SACHER and SAC show the similar learning time, SACHER provides the better learning performance than SAC as seen from Figs. \ref{fig_mujoco_1}-\ref{fig5}.


Fig. \ref{fig_mujoco_1} shows the cumulative reward of evaluation episodes for SAC, DDPG and SACHER during the learning (training) phase in several robotics environments implemented in OpenAI Gym \cite{Brockman2016}, which use the MuJoCo \cite{Todorov2012} physics engine. Each algorithm is trained for $10,000$ episodes with four different instances, and each episode carries out $1,000$ environmental steps. The solid curve corresponds to the mean, and the shaded area corresponds to the minimum and maximum rewards in four trials. Fig. \ref {fig_mujoco_1} shows that in overall, SACHER outperforms SAC and DDPG in terms of the learning speed and performance. As for the learning speed, SACHER reaches steady-state faster than SAC and DDPG in Ant-v2, Hopper-v2 and HalfCheetah-v2.
 Regrading the learning performance, SACHER achieves the higher average rewards than those of SAC and DDPG for every episode in all environments.

	{The simulation results of Environment I are shown in Figs. \ref{fig1} and \ref{fig2}. The learning curves of Environment I with SAC(left), DDPG(middle) and SACHER(right) are shown in Fig. \ref{fig1}. The brown-colored solid line represents the cumulative reward $R$ in each episode, and the yellow-colored dotted line shows the average of the last $20$ rewards in each episode. From Fig. \ref{fig1}(left and middle), although the learning process reaches steady-state by SAC and DDPG, the final position of the navigation paths generated by SAC and DDPG are rarely included in the landing area. During $2,000$ episodes, SAC provides only $32$ successful paths, and  DDPG provides only $15$ successful paths reaching the landing area. On the other hand, in Fig. \ref{fig1}(right), the navigation path generated by SACHER frequently reaches the landing area after the steady-state learning process. During $2,000$ episodes, SACHER provides $667$ successful paths reaching the landing area.}
	
The simulation results of navigation and tracking control of the hexarotor UAV with SACHER for Environment I are shown in Fig. \ref{fig2}.
The (red-colored) dotted line is the ideal path without obstacles (obtained from the straightforward computation), for which the straightforward computation yields the cumulative reward $R^*=-28.7854$. The (blue-colored) solid line indicates the path of the hexarotor UAV that follows the optimal path generated by SACHER via the backstepping controller (see Fig. \ref{fig_NN}). From Fig. \ref{fig2}, we can see that SACHER generates the almost ideal path reaching the landing area, where the corresponding computed cumulative reward is $R \approx -29$ that coincides with the ideal cumulative reward above ($R^*=-28.7854$). 
	
\begin{figure}[t!]
	\begin{center}
	\includegraphics[width=8.5cm]{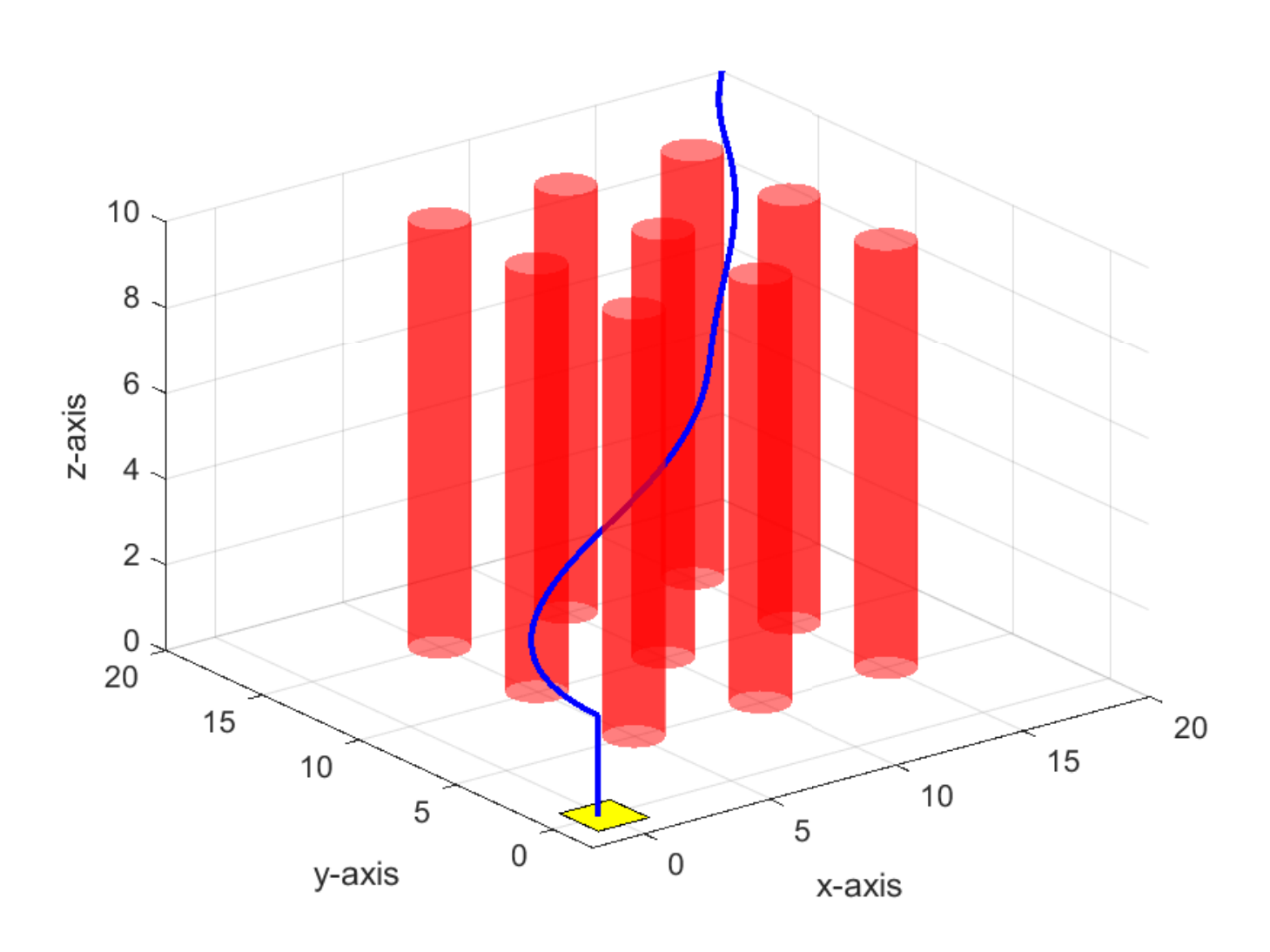}
	\includegraphics[width=8.5cm]{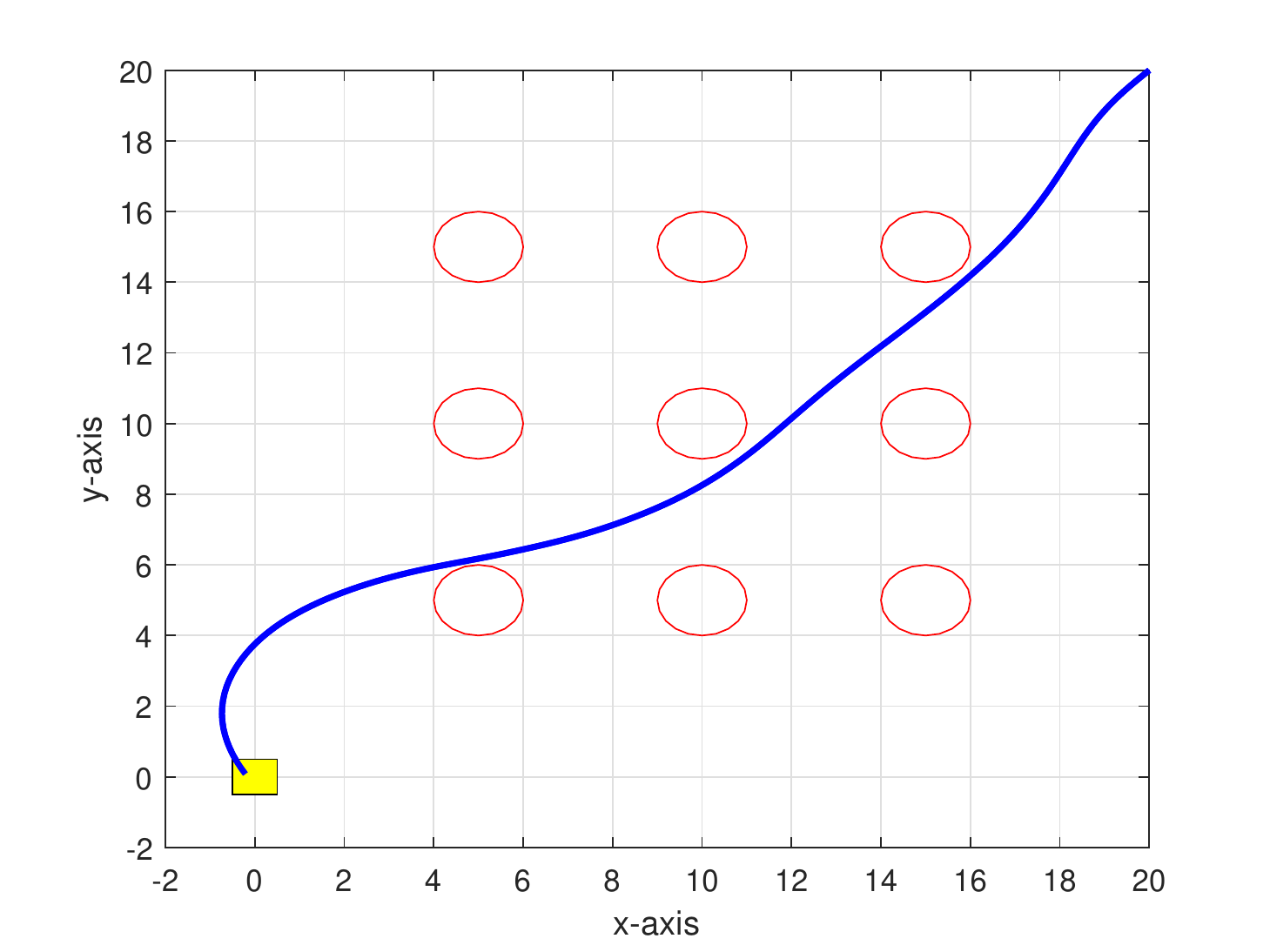}		
		\caption{\label{fig_R1} Environment II-A: The paths of the hexarotor UAV with SACHER under 9 obstacles when $v_2 = 8$.  }
	\end{center}
\end{figure}

\begin{figure}[t!]
	\begin{center}
	\includegraphics[width=8.5cm]{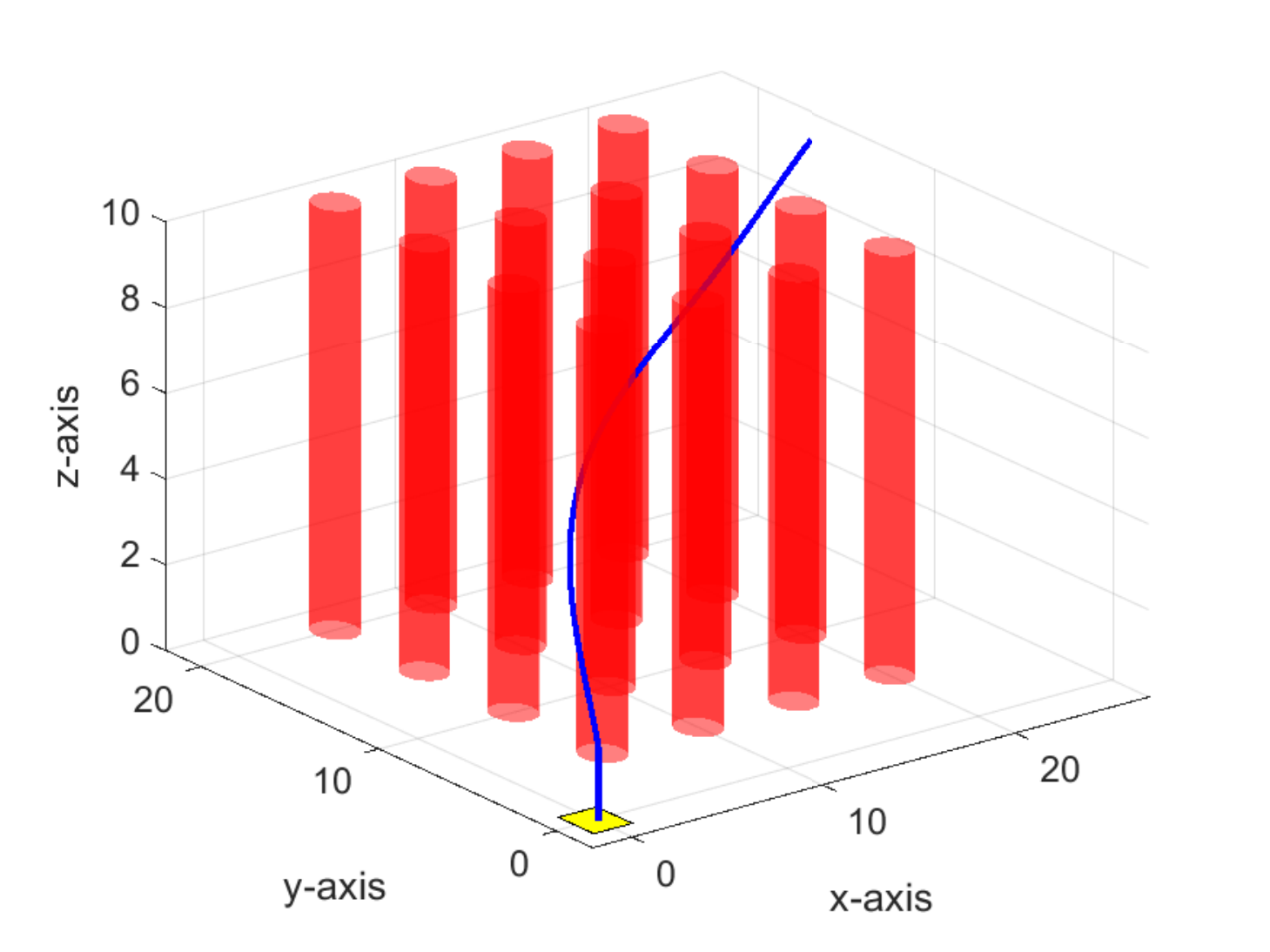}
	\includegraphics[width=8.5cm]{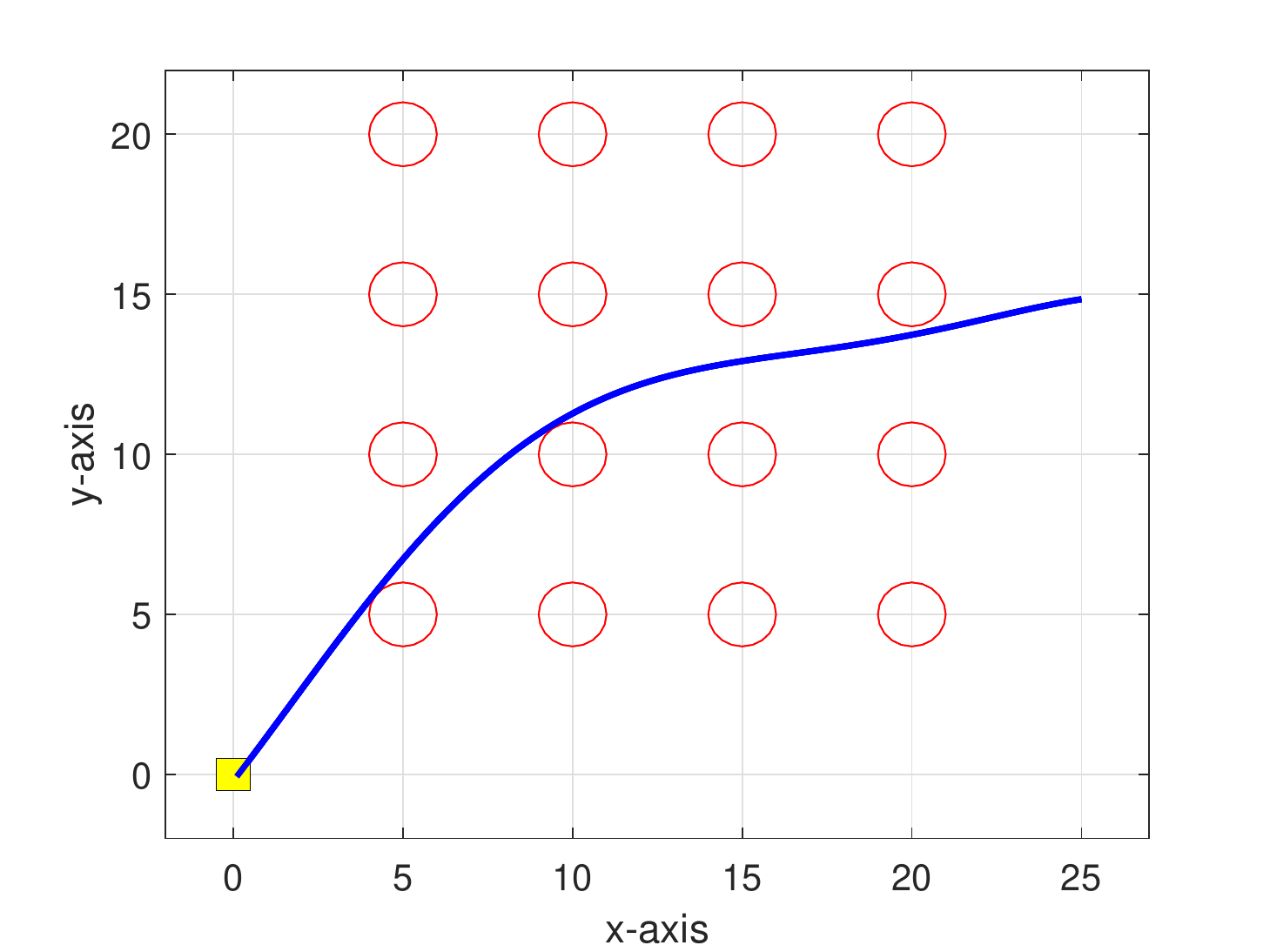}		
		\caption{\label{fig_R2} Environment II-B: The paths of the hexarotor UAV with SACHER under 16 obstacles. }
	\end{center}
\end{figure}	
	
The simulation results of Environment II are shown in Figs. \ref{fig3} and \ref{fig5}. The learning curves of Environment II with SAC(left), DDPG(middle) and SACHER(right) are shown in Fig. \ref{fig3}. The brown-colored solid line represents the cumulative reward $\bar{R}$ for each episode, and the yellow-colored dotted line shows the average of last $20$ rewards in each episode. From Fig. \ref{fig3}(left and middle), although the learning process reaches steady-state by SAC and DDPG, the final position of the navigation paths generated by SAC and DDPG are rarely included in the landing area. During $2,000$ episodes, SAC provides only $33$ successful paths, and DDPG provides only $13$ successful paths reaching the landing area. On the other hand, in Fig. \ref{fig3}(right), the navigation path generated by SACHER frequently reaches the landing area after the steady-state learning process. During $2,000$ episodes, SACHER provides $568$ successful paths reaching the landing area.

The simulation results of navigation and tracking control of the hexarotor UAV with SACHER for Environment II are shown in Fig. \ref{fig5}.
The paths of position and yaw angle, and their tracking errors are demonstrated in Fig. \ref{fig7}. The (red-colored) dotted line is the optimal navigation path generated by SACHER. The (blue-colored) solid line is the path of the hexarotor UAV controlled by the backstepping controller. From Fig. \ref{fig7}, it can be seen that the haxarotor UAV follows the optimal navigation path generated by SACHER with negligible tracking errors. In Fig. \ref{fig5}, the red-colored shapes are $9$ cylindrical obstacles, the turquoise-colored dashed line is the path of the hexarotor UAV that follows the navigation path generated by SAC, the green-colored dotted line is the path of the hexarotor UAV that follows the navigation path generated by DDPG, and the blue-colored solid line is the path of the hexarotor UAV that follows the optimal navigation path generated by SACHER. The navigation paths of SAC, DDPG and SACHER are the outcomes generated after the steady-state of the learning process. We observe that although the path of the hexarotor UAV with SAC and DDPG avoids the obstacles, the hexarotor UAV could not reach the landing area. Note that the path of the hexarotor UAV with SACHER successfully reaches the landing area while avoiding the obstacles. From our simulation results, we conclude that SACHER provides a more reliable path than SAC and DDPG with the cumulative reward of $\bar{R} \approx -31$ after completing the learning phase.
	
The simulation results of navigation and tracking control for the hexarotor UAV with SACHER for Environments II-A and II-B are shown in Figs. \ref{fig_R1} and \ref{fig_R2}, respectively. Note that Environments II-A and II-B are variations of Environment II, where
	\begin{itemize}
	\item Environment II-A modifies the velocity $v_1 = 8$ instead of $v_1 = 2$ in Environment II;
	\item Environment II-B considers $N= 16$ obstacles instead of $N= 9$ in Environment II.
	\end{itemize}
	From Figs. \ref{fig_R1} and \ref{fig_R2}, we can see that under the different settings in Environments II-A and II-B, the hexarotor UAV with SACHER successfully reaches the landing area while avoiding the obstacles. 
	
	Based on the above variations of Environment II, we may conclude that for any different environment settings, SACHER is able to generate the optimal navigation path for UAVs via Algorithm \ref{SACHER}. Hence, SACHER can be applied to any operations of UAVs with an appropriate environment design.

	\begin{remark}
	It should be mentioned that the learning curves in Figs. \ref{fig1} and \ref{fig3} show the reward, which do not affect the stability of the hexarotor UAV. Specifically, even if the reward is low in the learning curves in Figs. \ref{fig1} and \ref{fig3}, the trajectory of the hexarotor UAV generated by SACHER is smooth. However, in this case, the hexarotor UAV cannot reach the landing area. $\hfill{\square}$
	\end{remark}

	\section{Conclusions}\label{section5} 
In this paper, we have proposed a class of deep reinforcement learning algorithms, SACHER. In SACHER, HER improves the sample efficiency by allowing SAC to learn from both failures and successes when trying to achieve the goal. We have shown that SACHER achieves the desired optimal outcomes faster and more accurately than SAC. Our SACHER has been applied to the navigation and control problem of UAVs to generate the optimal navigation path for the UAV. Note that unlike the existing model-based approaches, SACHER in UAV navigation and control problems can be applied arbitrary models of UAVs, i.e., SACHER does not require specific information of UAVs. The effectiveness of SACHER has been validated through simulations, which include comparisons with state-of-the-art DRL algorithms, SAC and DDPG. One possible future work of this paper is to extend SACHER to the partially-observed case and apply it to the navigation and control problem of UAVs.

\bibliographystyle{IEEEtran}
\bibliography{ref.bib}

\end{document}